\documentclass[reprint,
preprintnumbers,
 amsmath,amssymb,
 aps,
pre
]{revtex4-2}

\usepackage{graphicx}
\usepackage{dcolumn}
\usepackage{bm}
\usepackage[document]{ragged2e}
\usepackage[mathlines]{lineno}
\usepackage{xfrac}
\usepackage{soul}
\usepackage{amsmath}
\usepackage{natbib}

\begin{document}

\setlength\parindent{15pt}

\title{Computational modeling of RNA-protein binding interactions under an external force}

\author{Danielle Wampler}
\affiliation{\textbf{Department of Physics}, The Ohio State University, 191 West Woodruff Ave, Columbus, OH 43210, USA.}
\author{Ralf Bundschuh}
\affiliation{\textbf{Department of Physics}, Department of Chemistry \& Biochemistry, Division of Internal Medicine, The Ohio State University, 191 West Woodruff Ave, Columbus, OH 43210, USA.\\}

\date{\today}

\begin{abstract}
RNA binding proteins play a crucial role in post-transcriptional gene regulation by controlling the transport, processing, and translation of their target RNAs. Post-transcriptional gene regulation leads to the differential expression of genetic material and loss of regulation or overregulation relates to a large range of cancers and diseases – many of which have directly been associated with RNA binding proteins and their target RNAs. To understand RNA, RNA binding proteins, and how they function in gene expression, it is essential to characterize how RNA binding proteins interact with their target RNAs. Here, we aim to assess the potential for single molecule force spectroscopy experiments to be used in the characterization of RNA-protein binding by investigating to what extent a change of extension due to RNA-protein binding is experimentally measurable and what aspects of the interaction can be deduced from such measurements. We predict the effect of protein binding on RNA force extension measurements via the open-source ViennaRNA package, which we have modified to simultaneously consider an external force, protein binding, and RNA secondary structure. From this work, we see protein concentration–dependent responses to external forces with discernable differences in predicted extensions around biologically relevant concentrations and a connection to protein binding domain geometry for several RNA binding proteins.
\end{abstract}

\maketitle

\pagebreak

\section{\label{Intro}Introduction\protect\\ }

RNA binding proteins together with their target RNAs are critical actors in many processes central to life. In post transcriptional gene regulation, the interactions between RNA and RNA binding proteins during the processing of pre-mRNA \cite{premRNAproc, premRNAcap, premRNAtail, premRNAsplice, RNAfashion}, RNA packaging and export \cite{RNAfashion}, RNA localization in the cell \cite{RNAloc}, mRNA translation \cite{RNAtrans}, and RNA decay \cite{RNAdechighway, RNAdecreg} play central roles in conferring biological functions. In nearly each of these post transcriptional gene regulation processes, disruptions to the interactions between RNA binding proteins and their target RNAs have been connected to various human cancers and diseases, such as spinal muscular atrophy, occulopharyngeal muscular dystrophy, fragile X syndrome, etc. \cite{PTGRdisease}. To better understand the specific roles that RNA binding proteins and their target RNAs have in differential gene expression, as well as to develop therapeutic approaches to the diseases that arise from disturbances to these interactions, it is essential that the interactions between RNA and RNA binding proteins be further characterized.

Owing to their importance in many central biological processes, there have been many techniques developed and implemented over the years to study RNA binding proteins and their interactions with RNA. Early approaches to identifying and characterizing RNA binding proteins involved in post transcriptional gene regulation involved classical biochemistry methods such as gel shifts. Predictive methods, which mostly focused on identifying domains in the protein structure known for interacting with RNA, or RNA binding domains, were also developed relatively early \cite{EvConRBP}. Later, many experimental approaches were developed to better characterize RNA binding protein roles and target RNAs. Cross linking experiments, such as cross linking and immunoprecipitation (CLIP) and its modifications, are a popular method of identifying sites of RNA bound to a given protein \cite{ApproachRev}. Alternatively, multiple experimental methods have been developed to identify proteins binding to a given RNA, which generally utilize cross-linking, RNA hybridization, mass spectrometry, and proximity labeling techniques \cite{ApproachRev}. Single molecule experimental techniques, often single molecule fluorescence, have also been applied in recent years to study RNA-protein interactions and offer a depth of insight into these interactions not available with ensemble approaches \cite{ApproachRev}. However, even with all of the recent work studying RNA-protein interactions, most of the known human RNA binding proteins, including their specific biological functions, still are not understood \cite{EvConRBP}. This gap in our understanding of crucial biological actors necessitates the development of new approaches and further experimentation. 

When RNA binding proteins interact with their target RNA, they often interact with the structures formed by the generally single stranded RNA molecules. Through these interactions with local RNA structures, RNA binding protein association affects the structures that its target RNA molecule is likely to adopt. Force spectroscopy approaches, such as magnetic and optical tweezers, are designed to measure end-to-end extensions of and external forces on a single, trapped target molecule \cite{OpticTweezerRef} and would therefore be appropriate to measure the changes in end-to-end extension of RNA molecules upon binding interactions with RNA binding proteins. Force spectroscopy methods are commonly used to extract distances to transition states, zero-force folding free energy barriers, and intrinsic folding rates \cite{pullingoutcomes} and have been readily applied to study RNA folding, protein folding, and protein-DNA interactions \cite{OpticTweezerRef}. While there have been some studies using optical tweezers to characterize RNA helicase activity \cite{helicase1, helicase2}, ribosome activity on RNA \cite{ribo1, ribo2, ribo3, ribo4}, and non-sequence specific RNA binding protein chaperone effects on RNA folding \cite{NCTAR}, the ability of force spectroscopy methods, such as magnetic and optical tweezers, to characterize the interactions between RNA and RNA binding proteins has not been leveraged yet to its full capacity. In this work, we will use a computational model to propose and highlight possible insights afforded by the application of single molecule force spectroscopy studies of sequence-specific RNA binding protein interactions with their target RNA.

The remainder of this manuscript is organized as follows. In Section~\ref{sec:ModelReview}, we first give a background on RNA structure and how it is modeled thermodynamically using the ViennaRNA program package \cite{ViennaRNA2.0}. Then, in Section~\ref{sec:InteracModel}, we build upon this RNA structure model with separate, established models of RNA structure under external force and RNA structure-protein interactions (in the absence of force) to obtain a computational model which describes pulling on RNA molecules while the RNA interacts with RNA binding protein. In Section~\ref{sec:Plausibility}, we show using this computational model that a force spectroscopy experiment involving pulling on an RNA actively interacting with protein would be capable of measuring the change in extension induced by protein-RNA binding at biologically relevant concentrations of free protein. Furthermore, in Sections~\ref{sec:Crossover}~and~\ref{sec:Geometry} we will use our computational model to highlight potential insights into binding interactions from this proposed experimental method and its significance. Finally, Section~\ref{sec:Discussion} summarizes our findings and provides suggestions and thoughts for future studies. 

\section{\label{sec:ModelReview}RNA Modeling Review\protect\\}
\subsection{\label{sec:SecStruc}RNA Secondary Structure\protect\\}

\begin{figure}
	\centering
	\includegraphics[width=0.5\columnwidth]{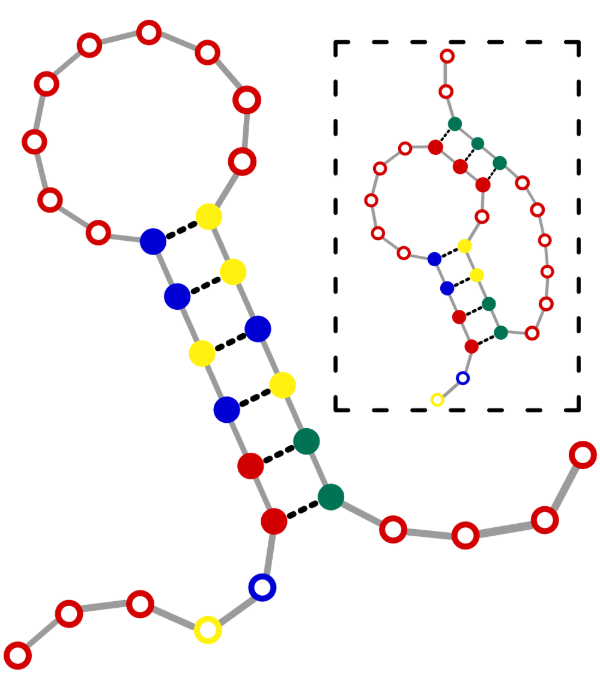}
	\caption{\label{RNAstruc} Example representation of RNA secondary structure. The four RNA bases A (green), U (red), C (blue), and G (yellow) are either base-paired (depicted with a filled center) or non-paired (unfilled center). The RNA backbone is depicted by a solid grey line  while the hydrogen bonds that form the base pairs are shown with dashed, black lines. In the inset, an example of a pseudoknot is shown. A pseudoknot is defined by the formation of base pairs between bases in a loop structure with bases outside of that helix-loop structure. Pseudoknot structures are explicitly excluded from this work.}
\end{figure}

RNA consists of the four nucleotide bases adenine~(A), uracil~(U), cytosine~(C), and guanine~(G). Owing to the propensity of these bases to interact with each other to form base pairs and to RNA's general existence as a single stranded polymer, RNA molecules can fold in on themselves to form secondary structures. RNA secondary structures are descriptions of an RNA molecule's base pairings, which can be either canonical (A-U, G-C) or Wobble (G-U) (Fig.~\ref{RNAstruc}). The secondary structures that RNA molecules can form in the model described here are limited to those where each base that it consists of is paired with at most one other base, where the bases in each base pair are at least $l_{min}=3$ nucleotides apart, and where there are no pseudoknots (Fig.~\ref{RNAstruc} inset). 

RNA secondary structure is an appropriate first approximation to describe RNA structure due to secondary structure interactions being generally stronger than tertiary interactions (three dimensional proximity interactions, including pseudoknots). The free energy of an RNA secondary structure is usually approximated according to the Turner energy model \cite{Turner2004} by adding up the free energies of each of its structural subunits (i.e. helices and loops), which are themselves determined from experimentally measured thermodynamic parameters. For clarity, we will describe methods in this text using a simplified energy model where the secondary structure energy $E_s$ is the sum of the energy contributions from each base pair $(k,l)$ (base $k$ paired with base $l$) that the structure $s$ consists of
\begin{equation}
	E_s = \sum_{(k,l) \in s} E(k,l),
\end{equation}
but we use the full Turner energy model in our computations.\\

\subsection{\label{sec:ModelStruc}Thermodynamic modeling of RNA secondary structure using ViennaRNA\protect\\}

RNA secondary structure is modeled in equilibrium by considering the ensemble of secondary structures $s$ available to a given sequence and the energies of these structures $E_{s}$. From this ensemble of structures and their energies $E_{s}$, the partition function
\begin{equation}
	Z = \sum_{s} e^{- \frac{E_s}{k_B T}}
\end{equation}
is obtained and from the partition function the ensemble free energy can be calculated as
\begin{equation}
	G=-k_{B} T \ln (Z).
\end{equation}

The number of structures available to a sequence grows exponentially with the length of the sequence $N$ \cite{Waterman}, making a simple sum over structures unreasonable for biologically relevant sequence lengths. Instead, the partition function $Z$ is calculated recursively \cite{McCaskill}. In this recursive calculation, initially the partition function over small segments of the RNA molecule is calculated, followed by the calculation of the partition function for progressively larger segments by piecing together the partition functions of the smaller segments it consists of until the total partition function for the entire given RNA molecule has been obtained.

\begin{figure}
	\centering
	\includegraphics[width=1.0\columnwidth]{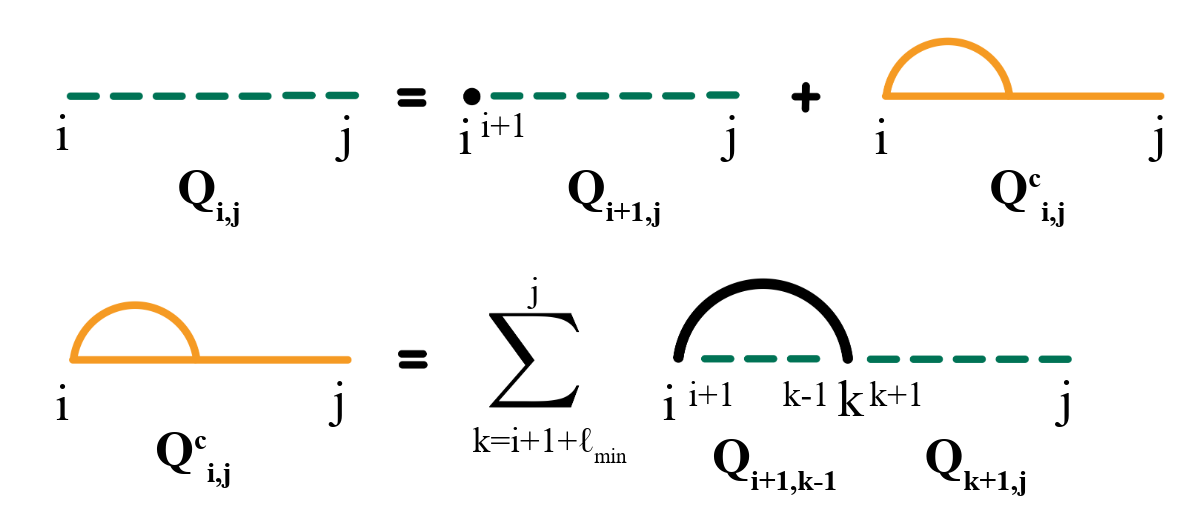}
	\caption{\label{simprec} Simplified representation of the recursive algorithm used to calculate the partition function of a given RNA molecule. The partition function for any given strand spanning from some base $i$ to base $j$, labeled $Q_{i,j}$, is the sum of the partition function over the structures where $i$ is unpaired, given by $Q_{i+1,j}$, and the constrained partition function over the structures where base $i$ is bound to some sufficiently separated base $k$, given by $Q^c_{i,j}$. RNA segments over which unconstrained partition functions are calculated are depicted with dashed, green lines, RNA segments over which constrained partition functions are calculated are depicted with solid, orange lines, unpaired bases are represented by a black dot, and base pairs are indicated by an arched black line.}
\end{figure}

A simplified representation of this partition function computation is shown in Fig.~\ref{simprec}. The partition function for an RNA segment spanning from base $i$ to base $j$ is given by $Q_{i,j}$ and is calculated according to the recursion relation 
\begin{equation}
	Q_{i,j} = Q_{i+1,j} + Q^c_{i,j}
\end{equation}
where $Q_{i+1,j}$ is the partition function over all of the structures where the first base from the left $i$ is unbound (which has already been calculated since it is shorter) and $Q^c_{i,j}$ is the constrained partition function over all the structures where the first base from the left $i$ is bound. This constrained partition function $Q^c_{i,j}$ is determined according to the recursive equation
\begin{equation}
	Q^c_{i,j} = \sum_{k=i+1+l_{min}}^{j} Q_{i+1,k-1} \cdot e^{-\frac{E(i,k)}{k_B T}} \cdot Q_{k+1,j}.
\end{equation}
In this recursive calculation of the constrained partition function $Q^c_{i,j}$, the sum is taken over all of the structures possible in the case of each base pair $(i,k)$ that the leftmost base $i$ could form, considering the minimum stretch of bases that must be in a hairpin loop $l_{min}$. When obtaining $Q^c_{i,j}$ for a given base pair case $(i,k)$, the constrained partition function calculation takes into consideration the partition function over all of the possible structures interior to this loop $Q_{i+1,k-1}$, the partition function over all of the possible bases exterior to (to the right of) this loop $Q_{k+1,j}$ (where by convention $Q_{j+1,j}=1$), and the partition function contribution of the base pair $e^{-\frac{E(i,k)}{k_B T}}$. In this recursion equation, $Q_{i+1,k-1}$ and $Q_{k+1,j}$ are already calculated partition functions.

A well established implementation of this recursive calculation of the partition function $Z$, but for the full Turner energy model, is that of the ViennaRNA package \cite{ ViennaRNA2.0}, which has been designed with a wide range of functionality with regards to RNA secondary structure prediction. Our approach here to model RNA secondary structure with both an external force and interacting RNA binding proteins is based on modifications to the ViennaRNA package and its recursive partition function calculations. For a more detailed description of the recursion relations used by the ViennaRNA package, refer to the work by Lorenz \emph{et al.} \cite{ViennaRNAConstraints}. \\

\section{\label{sec:InteracModel}RNA Interaction Modeling\protect\\}
\begin{figure}
	\centering
	\includegraphics[width=1.0\columnwidth]{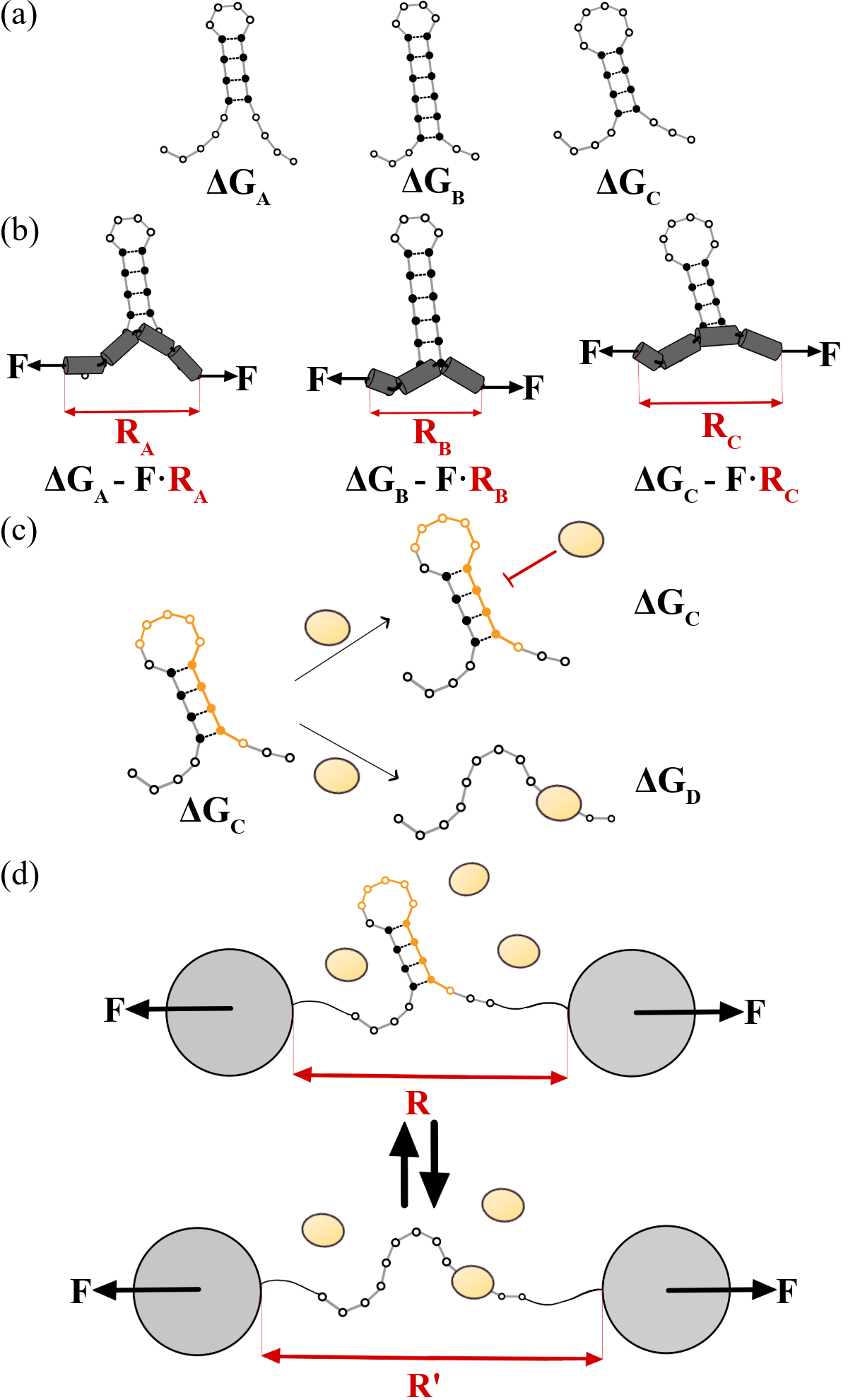}
	\caption{\label{overview}Overview of the development of this work's model, involving combining (a) RNA secondary structure modeling modified with (b) external force and with (c) interacting protein to get a (d) final model simulating applying an external force on an RNA molecule with interacting binding protein. (a)~Three example RNA secondary structures $s$ and their corresponding free energies $\Delta G_s$. (b)~The same three example RNA secondary structures $s$ under an external pulling force $F$. Each structure is superimposed with a depiction of the freely jointed chain model and their corresponding end-to-end extensions $R_s$ and modified free energies are included. (c)~An example RNA secondary structure with a protein binding site (highlighted in orange) is used to illustrate the competition between RNA secondary structure formation and RNA interactions with single-stranded RNA-binding proteins (yellow ovals). (d)~A force spectroscopy experiment that would be simulated by this final RNA-interaction model where competition between RNA secondary structure formation and interactions with single-stranded RNA-binding proteins occurs while an external force $F$ is applied on an RNA molecule trapped between two beads (grey circles) separated by extension $R$ (or $R'$). The binding footprint of the proteins (yellow ovals) depicted in (c) and (d) is 10 nucleotides in length.}
\end{figure}

Modeling of RNA secondary structure under an external force \cite{RNADenaturation} and in the presence of RNA binding proteins \cite{RBPBind} has been described before individually. Here, we first describe how we re-implement each concept individually in version 2.5.1 of the ViennaRNA package before discussing how we combine the two RNA interaction concepts (Fig.~\ref{overview}).\\

\subsection{\label{sec:ModelForce}Modeling RNA secondary structure under an external force\protect\\}

\begin{figure}
	\centering
	\includegraphics[width=1.0\columnwidth]{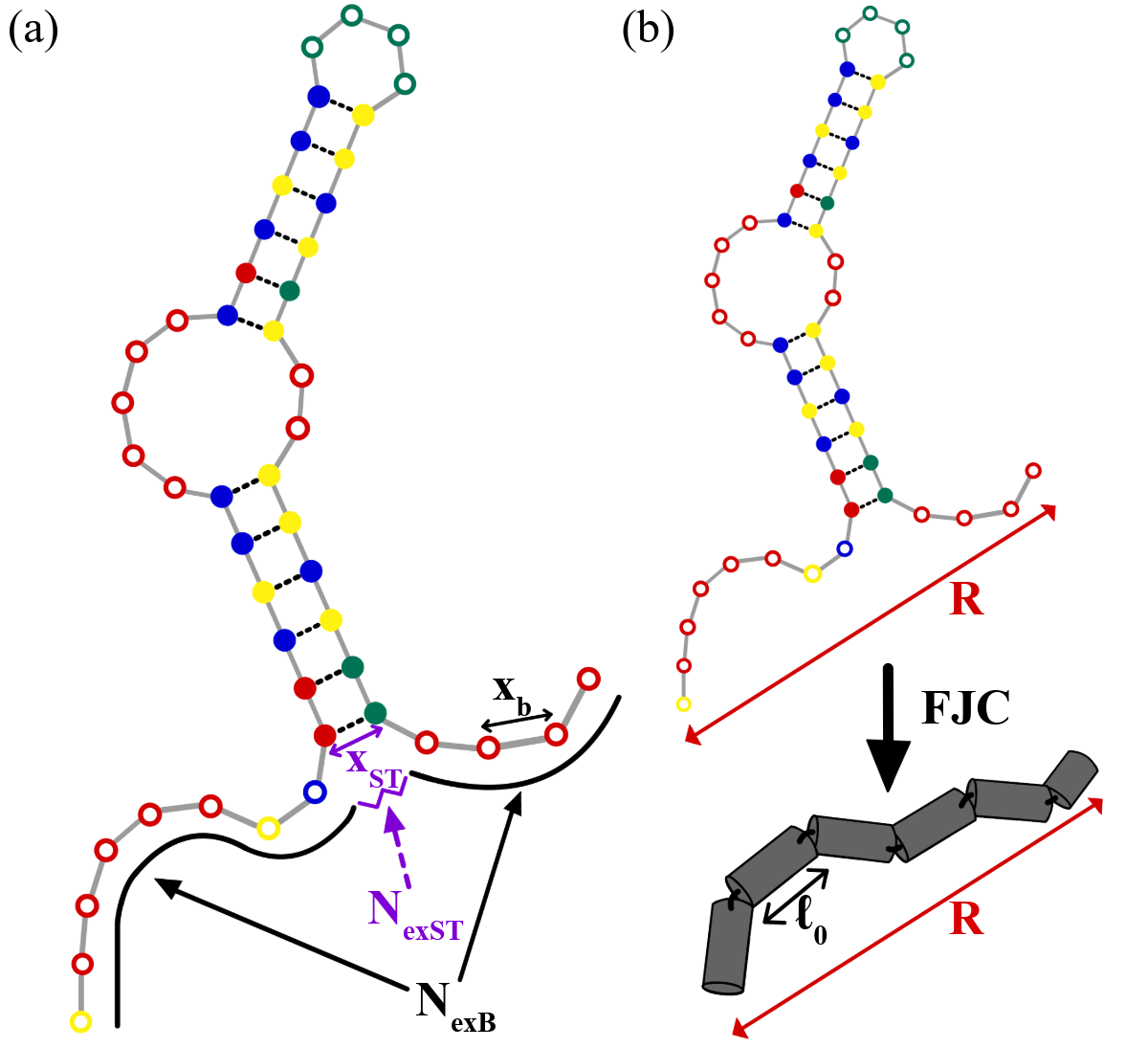}
	\caption{\label{paramsFJC} (a)~Definitions of relevant quantities for modeling an external pulling force on an RNA molecule. The external, non-paired bases are underlined by the solid lines and the external stem indicated by the underlining jagged line. The number of external, unpaired bases $N_{exB}$ and the number of external stems $N_{exST}$ are represented in this example as well as the distance between two subsequent single-stranded RNA bases $x_b$ and the extension of an external stem $x_{ST}$. (b)~In the freely jointed chain model, a polymer is represented by a chain of rods which are able to rotate completely independently of one another. The end-to-end extension of the RNA molecule $R$ is the same as the end-to-end extension of the representative chain and the length of an individual rod $\ell_0$ is the Kuhn length for single-stranded RNA.}
\end{figure}

Largely following the approach of Gerland \emph{et al.} \cite{RNADenaturation} to simulate pulling on RNA, we model the external part of an RNA molecule according to the freely jointed chain model. For some secondary structure $s$ formed by an RNA sequence of length $N$, we describe the external part of the RNA molecule in terms of the number of exterior bases $N_{exB,s}$ and the number of exterior stem structures $N_{exST,s}$ (Fig.~\ref{paramsFJC}(a)). In the freely jointed chain model, the series of exterior bases and external stem structures of the RNA are treated in space as a series of rods with length $\ell_0$ (Fig.~\ref{paramsFJC}(b)). The number of rods $N_{rod,s}$ in the freely jointed chain model of the backbone for a given structure is given by 
\begin{equation}
	N_{rod,s} = \frac{x_b \cdot N_{exB,s} + x_{ST} \cdot N_{exST,s}}{\ell_0}
\end{equation}
where $x_b$ is the distance between ssRNA bases and $x_{ST}$ is the distance along the backbone of ssRNA corresponding to an external stem (hairpin) structure.

In this model, each individual rod is allowed to rotate freely in space relative to the rod preceding it and each rod feels the external force $F$ being applied to the RNA molecule. In considering all of the possible states in space that the chain could rotate through, as well as the force applied to the RNA molecule, the partition function for each rod is found to be \cite{PhysBioCell}
\begin{equation}
Z_{rod} = \int e^{\frac{F \ell_0 \cos(\theta)}{k_B T}} d\Omega = \frac{k_B T}{\ell_0 F} \sinh(\frac{\ell_0 F}{k_B T}).
\end{equation}
From this, we arrive at a force dependent partition function 
\begin{equation}
Z_{total}(F) = \sum_{s} e^{- \frac{E_s}{k_B T}} [Z_{rod}(F)]^{N_{rod , s}}
\end{equation}
where again $N_{rod,s}$ is the number of rods and $E_s$ is the energy due to the secondary structure features for a given structure $s$. From this force-dependent partition function, the force-dependent ensemble free energy is given as 
\begin{equation}
G(F) = -k_B T \ln(Z_{total}(F)),
\end{equation}
which yields the force-extension curve
\begin{equation}
R(F) = - \frac{dG(F)}{dF}.
\end{equation}

In practice, we modify the ViennaRNA program to calculate the partition function $Z_{total}(F)$ and ensemble free energy $G(F)$ and we compute both for all forces $F$ in the range $ F \in [0~pN, 25~pN]$ in steps of $\Delta F = 0.02~pN$. Specifically, in the ViennaRNA program's implementation of the recursive calculation of the partition function $Z$ described in Sec.~\ref{sec:ModelStruc}, the exterior branch of the RNA is looped through separately from the interior loops and we can track the cases in which exterior unpaired stretches of a length $n$ are accounted for, where exterior stems are formed, and where a single exterior base is added. To modify the ViennaRNA program to compute $Z_{total}(F)$, the recursion equations are modified with additional factors of 
\begin{equation}
	Q_{mod,n}(F) = [Z_{rod}(F)]^{\frac{x_b \cdot n}{\ell_0}},
\end{equation} 
\begin{equation}
	Q_{mod,ST}(F) = [Z_{rod}(F)]^{\frac{x_{ST}}{\ell_0}},
\end{equation}
and 
\begin{equation}
	Q_{mod,B}(F) = [Z_{rod}(F)]^{\frac{x_b}{\ell_0}}
\end{equation}
in each of these respective cases. From these ensemble free energy values at each force, the force-extension curve $R(F)$ is determined as the numerical derivative using a central derivative
\begin{equation}
R(F) = - \frac{G(F + \Delta F) - G(F- \Delta F)}{2 \Delta F}.
\end{equation}

\subsection{\label{sec:ValForce}Validation of ssRNA pulling model\protect\\}

\begin{figure}
	\centering
	\includegraphics[width=1.0\columnwidth]{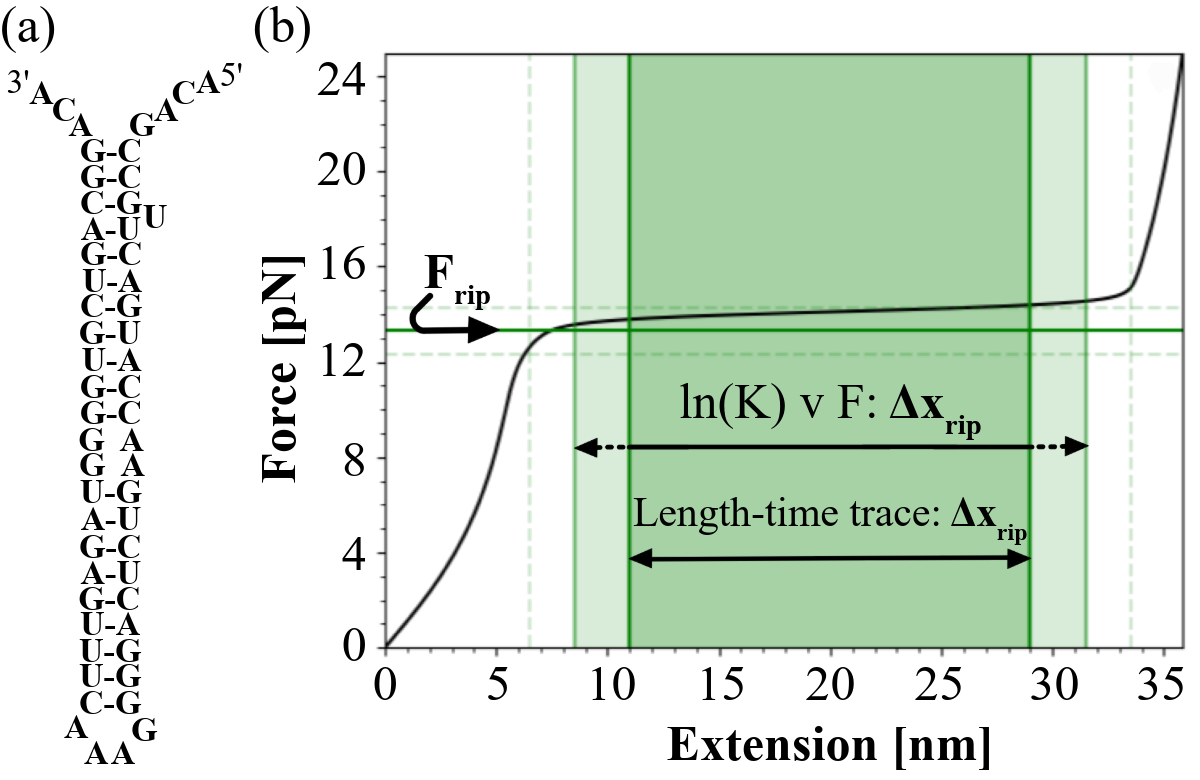}
	\caption{\label{exppullval}(a) The p5ab RNA hairpin sequence and expected secondary structure \cite{P5abExp}. (b) Simulated force-extension curve obtained from our RNA structure model with external force modifications (black curve) compared to the rupture force $F_{rip}$ and change in extension $\Delta x_{rip}$ measurements made by Liphardt \emph{et al.} for p5ab in the absence of Mg$^{2+}$ \cite{P5abExp}. The $F_{rip}$~$=$~$13.3$~$\pm$~$1$ pN rupture force measured experimentally is indicated by the horizontal solid green line (and the experimental error indicated by the horizontal dashed green lines). The rip extension $\Delta x_{rip}$~$=$~$18$ nm measured from constant force measurements is indicated by the vertical solid darker green lines and corresponding enclosed shading and the rip extension $\Delta x_{rip}$~$=$~$23$~$\pm$~$4$ nm extracted from $\ln$(k) vs. F plots is indicated by the vertical solid lighter green lines (and lighter shading). The experimental error on the rip extension (as extracted from the $\ln$(k) vs. F plots) is shown with the vertical dashed green lines. The experimental measurements for the rip extensions $\Delta x_{rip}$ shown on the force extension plot have been centered on the simulated rip to highlight the agreement between the simulation with experiment \cite{P5abExp}.}
\end{figure}

We validated our implementation of this RNA pulling model by simulating a poly-U RNA force extension curve and comparing it to the analytically known FJC force extension curve as well as by comparing simulations of pulling on the p5ab RNA hairpin (Fig.~\ref{exppullval}(a)) to the relevant experimental measurements made by Liphardt \emph{et al.} \cite{P5abExp} (Fig.~\ref{exppullval}(b)). In the poly-U pulling simulation comparison to the analytical FJC force extension curve, we see complete agreement between the simulation and the analytical solution. 

In the validation of the p5ab hairpin pulling simulation, we compare our prediction to the experimental results via their measurements of the force at which the hairpin rips $F_{rip}$ and the change in extension of the p5ab RNA over the ripping process $\Delta x_{rip}$. In the absence of Mg$^{2+}$, the ripping force was measured by Liphardt \emph{et al.} to be $F_{rip} = 13.3 \pm 1$ pN from force extension curve measurements. In the same study, there were two measurements made of the change in extension of the p5ab RNA over the rip: $\Delta x_{rip} = 18 \pm 2$ nm from constant force measurements and $\Delta x_{rip} = 23 \pm 4$ nm using the slope from a plot of $\ln$(k) vs. F using an analogous van't Hoff formula \cite{P5abExp}. 
Displayed in Fig.~\ref{exppullval}(b), we see convincing agreement between p5ab pulling simulations and experimental measurements of $F_{rip}$ and $\Delta x_{rip}$ for p5ab within experimental error.\\

\subsection{\label{sec:ModelProt}Thermodynamic modeling of protein-RNA binding in the absence of external force\protect\\}

\begin{figure}
	\centering
	\includegraphics[width=1.0\columnwidth]{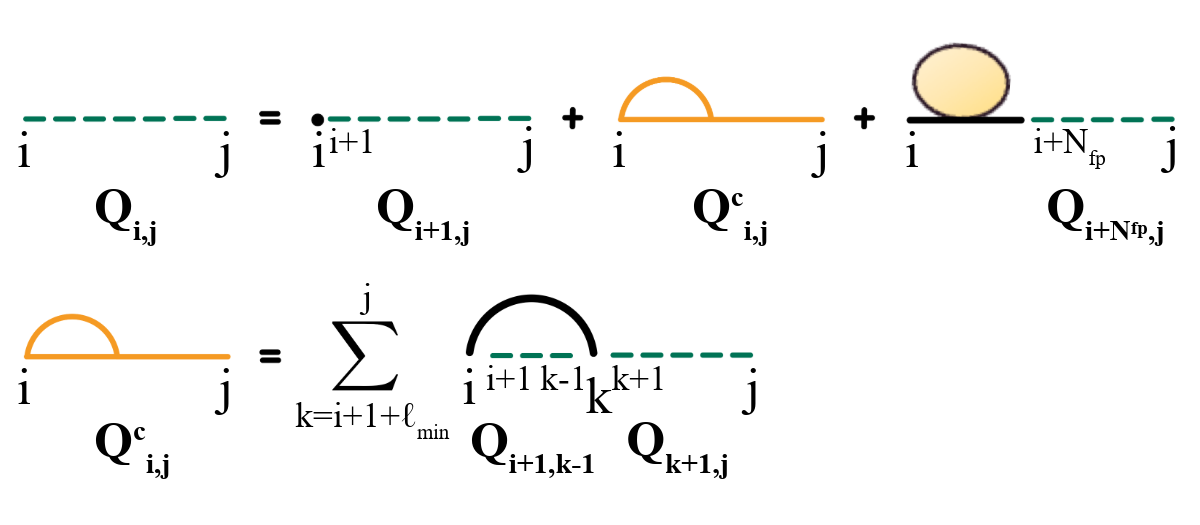}
	\caption{\label{finpartfunc} The recursive algorithm used to calculate the partition function of a given RNA molecule in the presence of a single-stranded RNA binding ligand. The original recursion equation used to calculate the partition function $Q_{i,j}$ for any given strand spanning from some base $i$ to base $j$ in the given RNA sequence is modified to also include the partition function over the structures where the substrand starting at base $i$ with length matching the size of the ligand's binding footprint $N_{fp}$ is bound by the protein. The bound unstructured domain in this partition function representation is depicted with a straight solid black line (with an interacting protein shown by the associated yellow oval).}
\end{figure}

To account for single stranded RNA binding proteins in the absence of external force, we leverage the ViennaRNA package's ability to allow for ligand binding to ``unstructured domains" (unpaired RNA strands or segments) in the recursive calculation of the partition function. More specifically, as ViennaRNA recursively iterates through the allowed structures of the given RNA sequence, the recursion additionally accounts for the structures with the RNA binding protein bound to unstructured RNA regions of length equal to that of the binding footprint of the protein $N_{fp}$ (Fig.~\ref{finpartfunc}). 
The partition function for these structures with $M$ proteins bound are corrected by a factor of $\prod_{m=1}^M(\frac{c}{{K_D}_m})$ according to the free protein concentration $c$ and to the absolute $N_{fp}$-mer binding affinities $\left\{{K_D}_m\right\}^{-1}$ for the protein of interest at each of the corresponding bound motifs $m$. The $N_{fp}$-mer binding affinities incorporated in this model describe the binding affinities of the protein to each possible binding site sequence of length of the protein's binding footprint $N_{fp}$. They were determined from relative RNAcompete \cite{RNAcompete} binding affinities that were calibrated using corresponding experimental binding affinity measurements \cite{RBPBind}. From this model, we are able to predict concentration-dependent partition functions $Z(c)$ and thus concentration-dependent free energies $G(c)$.\\

\subsection{\label{sec:ValProt}Validation of protein-RNA binding model\protect\\}

We validate the protein-RNA binding interaction modeling by predicting effective binding constants $K_D$ for a set of RNA binding proteins and corresponding sequences for which experimental measurements of the effective binding constants have been made. To this end we simulate RNA interactions with four single-stranded RNA binding proteins: HuR, RBFOX1, U2AF2, and KHDRBS3. The choice of these proteins is due to their absolute binding affinity data availability \cite{RBPBind}. HuR is a widely studied regulator of mRNA that binds U and AU-rich motifs in untranslated regions (UTR) of target mRNAs \cite{HuRBackground}. RBFOX1 is a neuron-specific effector of alternative splicing \cite{RBFOX1Background} that binds majorly in highly conserved intronic regions in their target mRNA \cite{RBFOX1Background2}. U2AF2 (or U2AF 65) is involved in pre-mRNA splicing and the 3' end processing of pre-mRNA by associating primarily with polyprimidine tracts in intronic regions of target mRNAs \cite{U2AF2Background}. KH RNA-binding domain containing, signal transduction associated 3 RNA-binding protein KHDRBS3 (also known as T-STAR \cite{TSTAR}) is a testis and brain-tissue specific regulator of splicing \cite{KHDRBS3}. Effective binding constants can be predicted using ViennaRNA by numerically determining at what concentration $c'$ the binding probability
\begin{equation}
	P_{bound}(c') = 1 - e^{-\frac{G(c=0) - G(c=c')}{k_B T}}
\end{equation}
becomes $P_{bound}(c' = K_D) = 0.5$ $(\pm 0.0001)$. These effective binding constant predictions are compared both against similar predictions by the previously published RBPBind software \cite{RBPBind} and against corresponding experimental effective binding constant measurements. For this comparison, experimental binding affinity data on 13 HuR-interacting \cite{HuRExpSeq}, 6 U2AF2-interacting \cite{U2AF2ExpSeq}, 8 RBFOX1-interacting \cite{RBFOX1ExpSeq}, and 7 KHDRBS3-interacting \cite{KHDRBS3ExpSeq} short sequences was used.
\begin{figure}
	\centering
	\includegraphics[width=1.0\columnwidth]{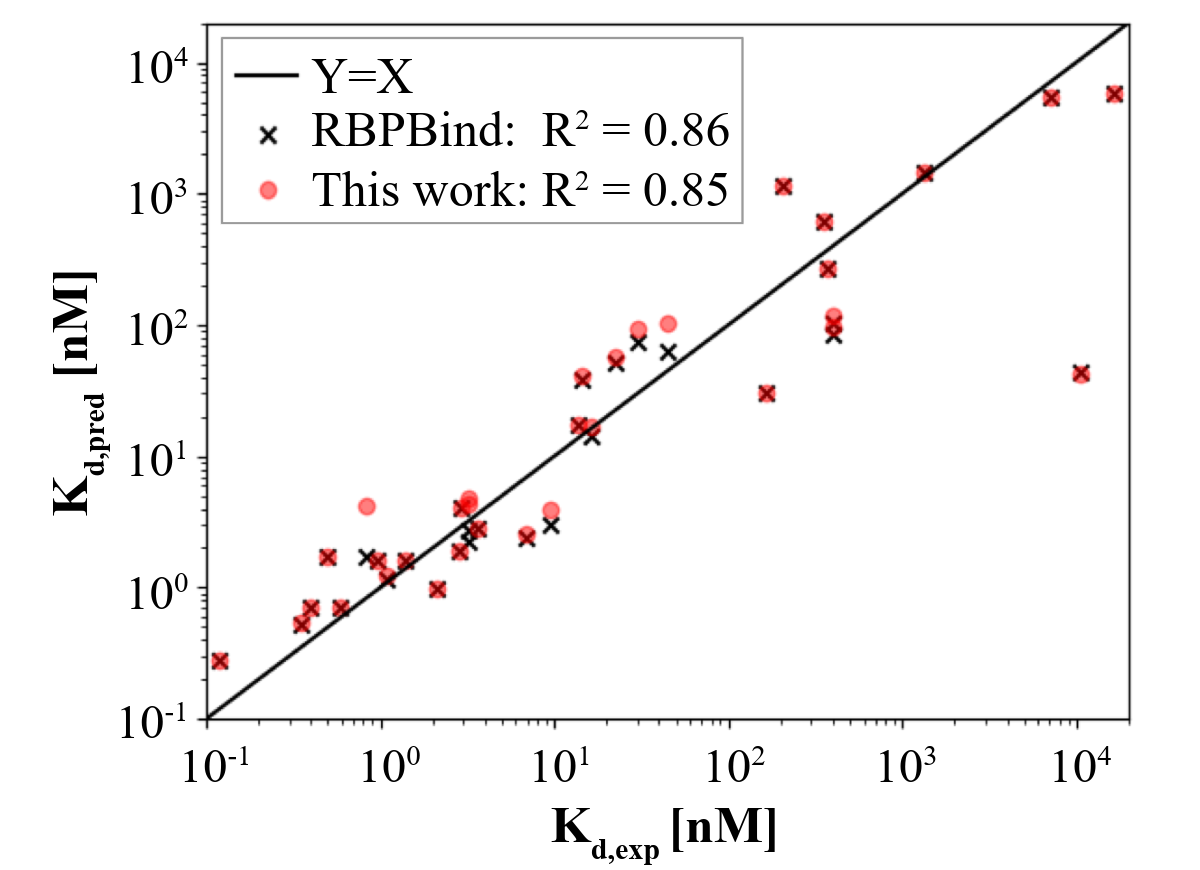}
	\caption{\label{protval} Comparison of effective binding constants predicted by this work's RNA-protein interaction model and by the similar model RBPBind \cite{RBPBind} to experimental measurements of corresponding effective binding constants for a set of short sequences interacting with relevant proteins HuR, RBFOX1, U2AF2, and KHDRBS3 \cite{HuRExpSeq, RBFOX1ExpSeq, U2AF2ExpSeq, KHDRBS3ExpSeq}. Proximity of data points to the diagonal solid black line indicate agreement between prediction and experimental findings of binding constants and the large $R^2$ values found for both this work and for the RBPBind prediction indicate strong positive correlation between prediction of effective protein binding constants and their experimental values.}
\end{figure} 
Very strong correlation ($R^2 = 0.99$) was found between our predictions and those made by RBPbind, as would be expected given that RBPBind uses the same modeling approach integrated into an earlier version of ViennaRNA that did not yet support ligand binding; strong correlation ($R^2 = 0.85$) was also found between the model's predictions and experimental measurements (Fig.~\ref{protval}).\\

\subsection{\label{sec:ModelAll}Thermodynamic modeling of protein-RNA binding under external force\protect\\}

Finally, to simulate RNA under external force in the presence of protein, we combine the described models for pulling on RNA and for protein binding and take into additional consideration that any bound external RNA would be under force. We model external protein bound sites as a single rod of length corresponding to the protein binding domain length $\ell_p$ instead of multiple rods of length $\ell_0$. 

\begin{figure}
	\centering
	\includegraphics[width=0.5\columnwidth]{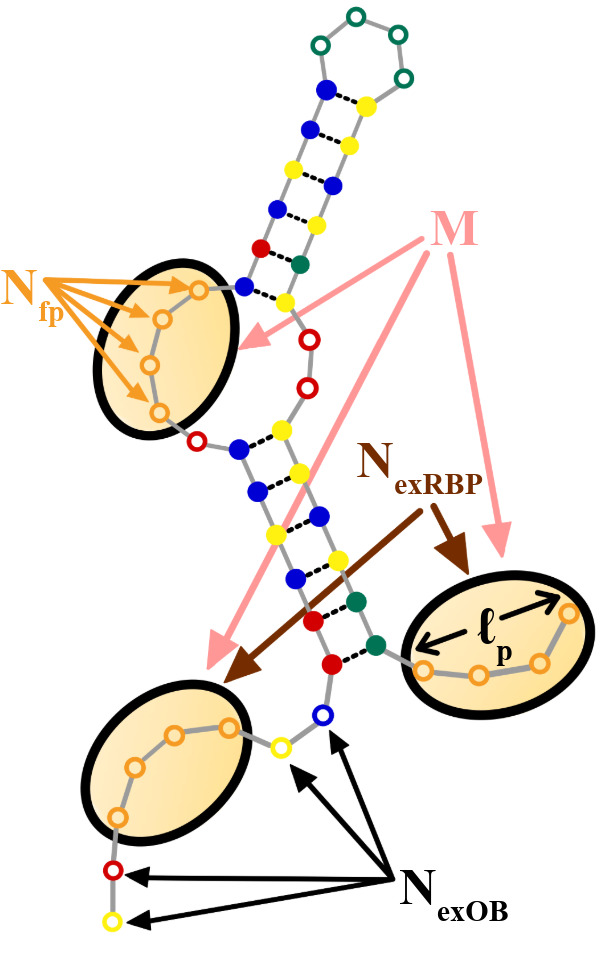}
	\caption{\label{paramsFJCwP} Definitions of relevant quantities for modeling an external pulling force on an RNA molecule with interacting single-stranded RNA-binding protein. The number of external, unbound and unpaired bases $N_{exOB}$, the size of the protein binding footprint $N_{fp}$, the protein binding domain geometry (length) $\ell_p$, the total number of bound proteins $M$, and the number of externally bound proteins $N_{exRBP}$ are shown.}
\end{figure}

\begin{figure}
	\centering
	\includegraphics[width=1.0\columnwidth]{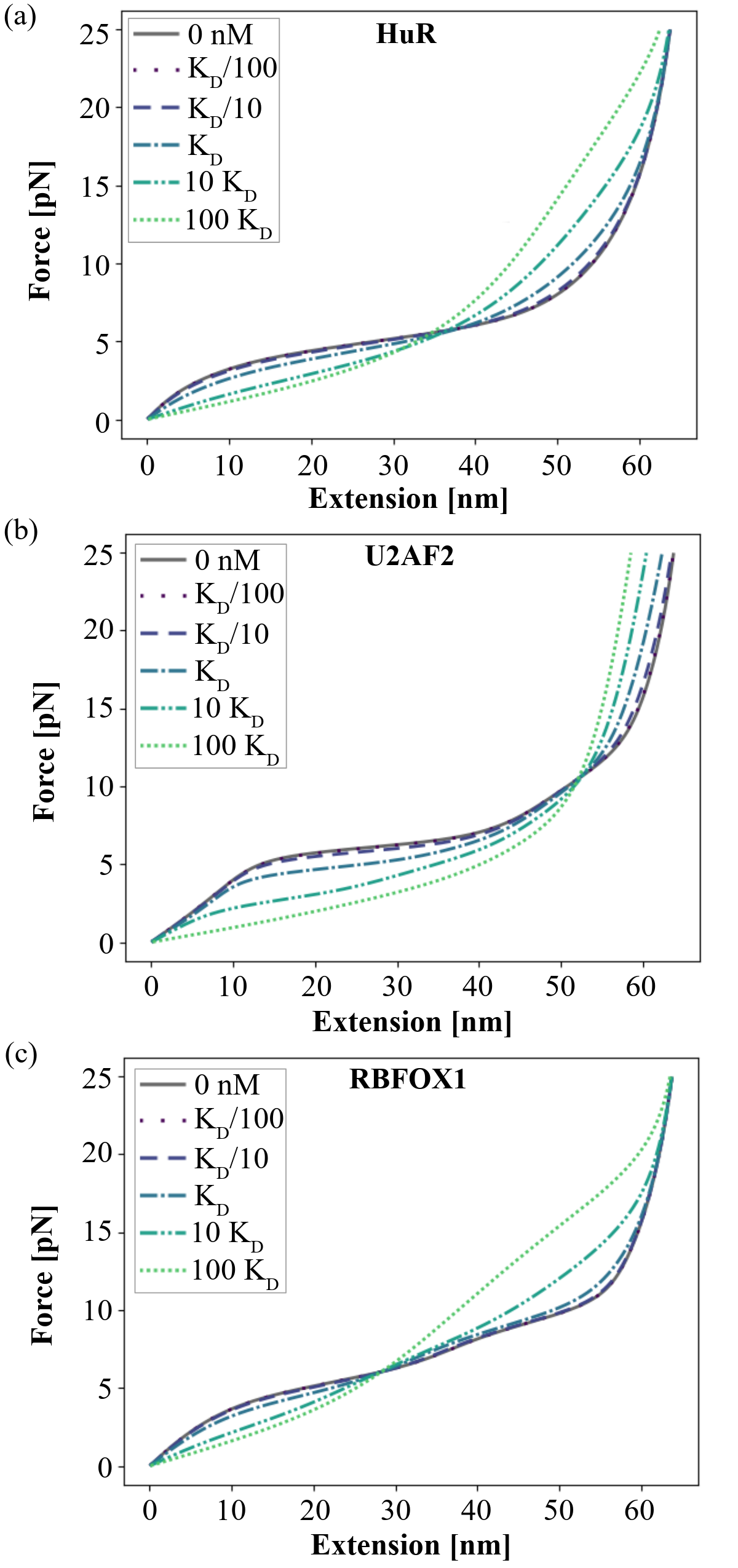}
	\caption{\label{FECtool} Concentration-dependent force extension simulations for three example physiological RNA sequences with naturally occurring binding sites of (a) HuR, (b) U2AF2, or (c) RBFOX1. The six force extension curves shown in each of the example plots correspond to varying simulated free concentration of the relevant protein, including a simulation with no protein. In each case, the predicted effective binding constant $K_D$ for the given RNA sequence and the protein it is interacting with (HuR, U2AF2, or RBFOX1) is used as a reference point to increment the free protein concentration being simulated.}
\end{figure}

Specifically, in ViennaRNA's recursive calculation of the partition function in the exterior branch of the RNA, we can track when a protein is externally bound and modify the partition function to include a factor of 
\begin{equation}
	Q_{mod,P}(F) = \frac{k_B T}{\ell_p F} \sinh(\frac{\ell_p F}{k_B T})
\end{equation}
rather than the partition function correction $Q_{mod, n=N_{fp}}(F)$ (Sec.~\ref{sec:ModelForce}) that would have been applied if the stretch of $N_{fp}$ bases were unbound. With this, we can calculate partition functions given an RNA sequence, force $F$, and free concentration $c$ of protein: 
\begin{equation}
	\begin{aligned}
		Z(F,c) &= \sum_{s}\left\{ W_{structure}\left[W_{unboundRNA}(F)\right] \right. \\
		&	 \left. \left[W_{binding}(c)\right] \left[W_{boundRNA}(F)\right] \right\} \\
		&= \sum_{s} \left\{ e^{- \frac{E_s}{k_B T}} \left[\frac{k_B T}{\ell_0 F} \sinh\left(\frac{\ell_0 F}{k_B T}\right)\right] ^{\frac{x_{exRNA, s}}{\ell_0}} \right. \\
		&	 \left. \left[\prod_{m=1}^{M_s}(\frac{c}{K_{D,m}})\right] \left[\frac{k_B T}{\ell_p F} \sinh\left(\frac{\ell_p F}{k_B T}\right)\right]^{N_{exRBP, s}} \right\}
	\end{aligned}
\end{equation}
where, for a structure $s$, $x_{exRNA}$ given by
\begin{equation}
	x_{exRNA,s} = x_b \cdot N_{exOB, s} + x_{ST} \cdot N_{exST,s}
\end{equation}
is the extension due to the unbound external RNA and exterior RNA stem structures, $N_{exOB,s}$ is the number of open (unbound) external bases, $M_s$ is the number of bound $N_{fp}$-mer motifs, $K_{D,m}$ is the binding affinity for the bound motif $m$, and $N_{exRBP,s}$ is the number of externally bound RNA binding proteins (Fig.~\ref{paramsFJCwP}). 
From the partition function $Z(F,c)$, we can obtain a prediction of $G_c(F)$ and subsequently $R_c(F)$ using the same central derivative method described previously (Sec.~\ref{sec:ModelForce}). With our prediction of $R_c(F)$, we are able to predict concentration-dependent force-extension curves for a given RNA sequence and RNA binding protein (Fig.~\ref{FECtool}).\\

\section{\label{sec:Results}Results\protect\\ }

\subsection{\label{sec:Plausibility}Proposed RNA force spectroscopy with interacting proteins is experimentally plausible\protect\\ }

The primary aim now that we are able to predict concentration-dependent force extension curves is to assess the capability of these simulated RNA force spectroscopy experiments that we propose here to measure RNA-protein binding interactions. To this end and for the remainder of this study, we will simulate RNA interactions with three single-stranded RNA binding proteins: HuR, RBFOX1, and U2AF2. The continued choice of these three proteins, and the henceforth exclusion of KHDRBS3, is due to the availability (and lack thereof, in the case of KHDRBS3) of information on their natural binding sites, which allow for the determination of physiologically relevant RNA sequences that will be used for simulation purposes. For each of these three proteins, using their experimentally determined natural binding sites \cite{HuRData, RBFOX1Data, U2AF2Data}, we obtain 100 physiological sequences 100 nucleotides in length (Appendix~\ref{sec:Seqs}). With these three RNA binding proteins, and their 100 physiological sequences each, we assess whether protein binding is measurable with concentration-dependent force extension curves.

\begin{figure}
	\centering
	\includegraphics[width=1.0\columnwidth]{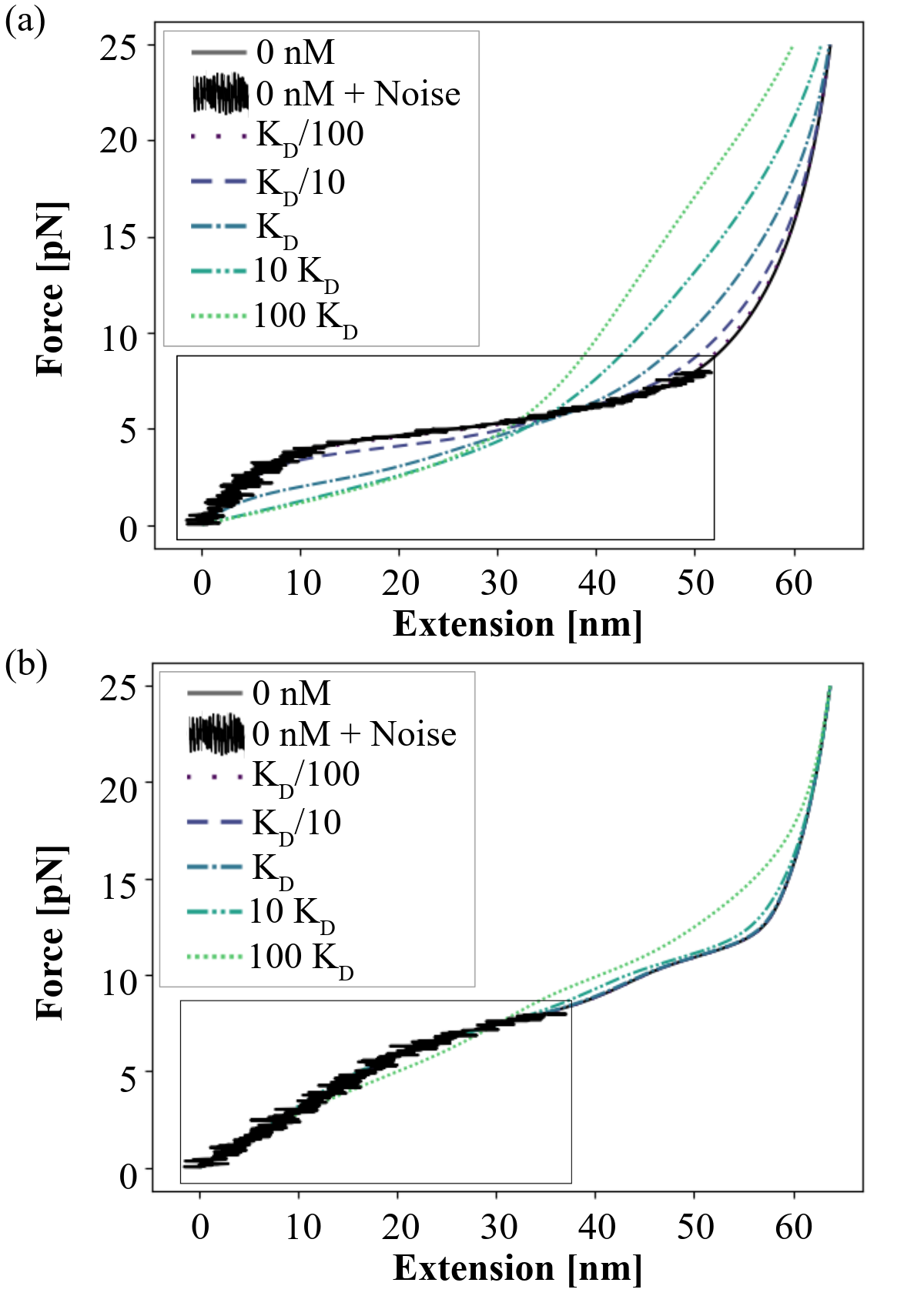}
	\caption{\label{distinex} Example concentration-dependent force extension curves where protein binding is (a) distinguishable and (b) indistinguishable. Simulated gaussian noise on the zero protein force extension curve is shown on top of the force extension curves without protein (0 nM) in both example plots. The boxes in either plot approximately indicate the range over which protein binding distinguishability is evaluated. (a) Protein binding is clearly distinguishable as the force extension curves at concentrations of $K_D$ and larger can clearly be seen to deviate from the curve without protein. (b) Protein binding appears indistinguishable as in the highlighted region none of the depicted force extension curves with interacting protein deviate significantly from the simulated gaussian noise on the zero protein force extension curve.}
\end{figure}

\begin{figure}
	\centering
	\includegraphics[width=1.0\columnwidth]{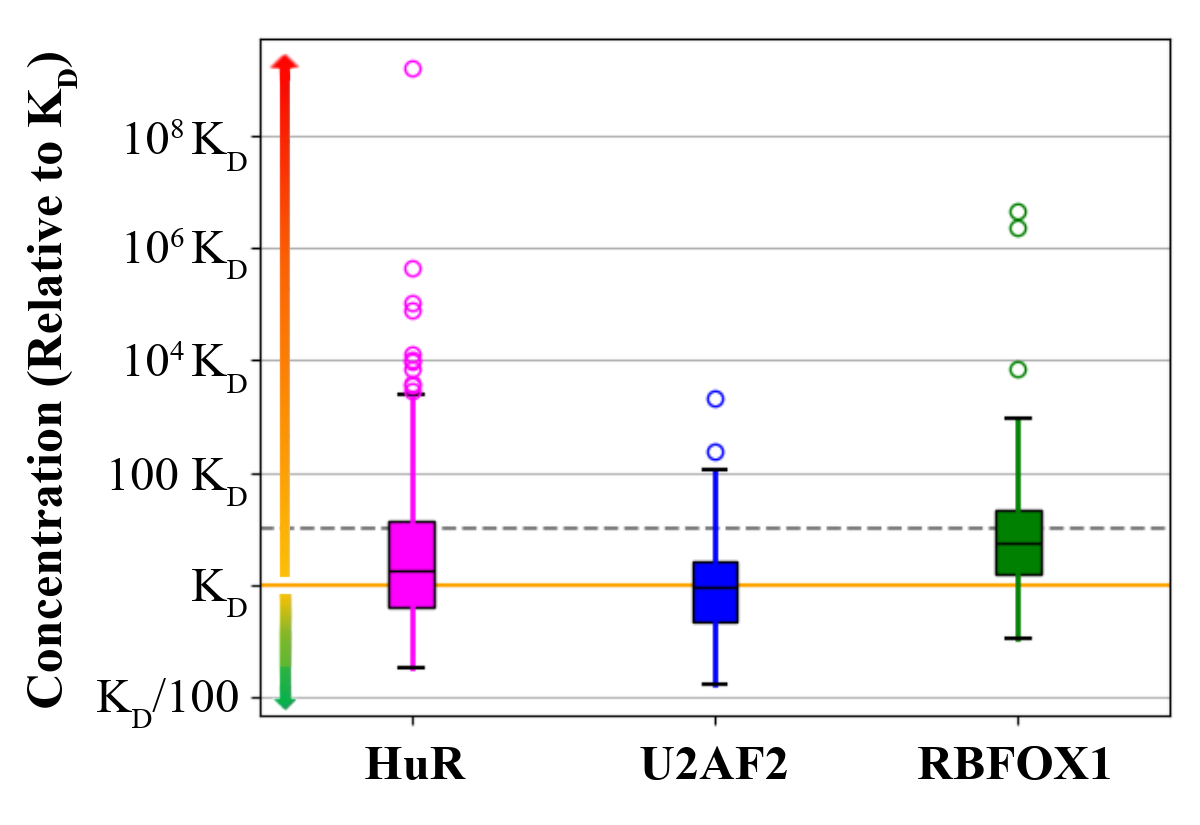}
	\caption{\label{firstdistin} Distributions of free protein concentrations at which protein binding first becomes distinguishable for HuR, U2AF2, and RBFOX1. Concentrations are reported relative to the corresponding predicted effective binding constant $K_D$ (horizontal solid yellow line) for the corresponding protein and molecule to give a normalized measure of biological relevance. Qualitatively, the degree of biological relevance is indicated by the vertical gradient arrows on the left edge of the plot, where concentrations when first distinguishable are considered more biologically relevant if they are within an order of magnitude of, or any amount less than, their predicted effective binding constant $K_D$ (yellow and green vertical gradient regions, below horizontal grey dashed line). First distinguishable protein concentrations above an order of magnitude larger than the corresponding predicted effective binding constants $K_D$ are considered less biologically relevant (orange and red vertical gradient regions, above horizontal grey dashed line).}
\end{figure}

In order to determine whether protein binding could be measurable from experimental concentration-dependent force extension curves, we simulate gaussian noise on the zero-protein force extension curves and determine if and at what concentration force extension curves with protein binding become distinguishable from the noisy zero-protein force extension curves. After calculating force-extension data given an RNA sequence, noise is generated by adding deviations to the predicted extension at each force independently drawn from a normal distribution with a standard deviation of 1 nm, corresponding to the approximate 1 nm precision of optical trap measurements \cite{P5abExp}. To determine the protein concentration at which protein binding first becomes distinguishable from the simulated noise, we incrementally increase the modeled protein concentration until the predicted force extension curve deviates significantly from the simulated noisy zero-protein force extension curve. Specifically, for each sequence, in the absence of protein, the areas under 100 noisy force extension curves up to a maximum force of 7.0 pN were collected  and their average noisy area and standard deviation calculated. Protein binding is deemed distinguishable at a given concentration if the area under the corresponding force extension curve deviates by more than 2 standard deviations from the average noisy area (Fig.~\ref{distinex}, Appendix~\ref{sec:FirstDistin}). From this, we find that for all but one of the 300 RNA molecules across the 3 proteins that we simulate pulling for, protein binding eventually becomes distinguishable.
Additionally relevant to the question of the practicality of this simulated experiment is the biological relevance of the concentrations at which protein binding becomes distinguishable. We use predicted effective binding constants $K_D$ at zero force for each sequence and their corresponding protein as a benchmark for biological relevance. That is, for a given protein and sequence, if the concentration at which protein binding becomes distinguishable from the simulated noise is within an order of magnitude of its corresponding predicted effective binding constant, or any amount smaller than it, then we consider this to be biologically relevant. The full distribution of concentrations (relative to their predicted effective binding constants) when binding becomes first distinguishable for each protein is shown in Fig.~\ref{firstdistin}. We find 74\% of the concentrations when protein binding first becomes observable are biologically relevant. Specifically, 70\% of HuR simulations, 91\% of U2AF2 simulations, and 62\% of RBFOX1 simulations exhibited biological relevance. Notably, 53\% of the U2AF2 simulated experiments and 39\% of the simulated HuR experiments showed that protein binding would be observable at concentrations even below their predicted effective binding constant.

We conclude from the near totality (99.7\%) of simulations where protein binding is distinguishable and from the distributions of concentrations where we first distinguished protein binding being centered at least within an order of magnitude of relevant predicted effective binding constants that these simulated RNA force spectroscopy experiments with interacting proteins are experimentally plausible.\\

\subsection{\label{sec:Crossover} Protein concentration-dependent force extension curves feature crossover points where extension becomes independent of protein concentration\protect\\ }

\begin{figure}
	\centering
	\includegraphics[width=1.0\columnwidth]{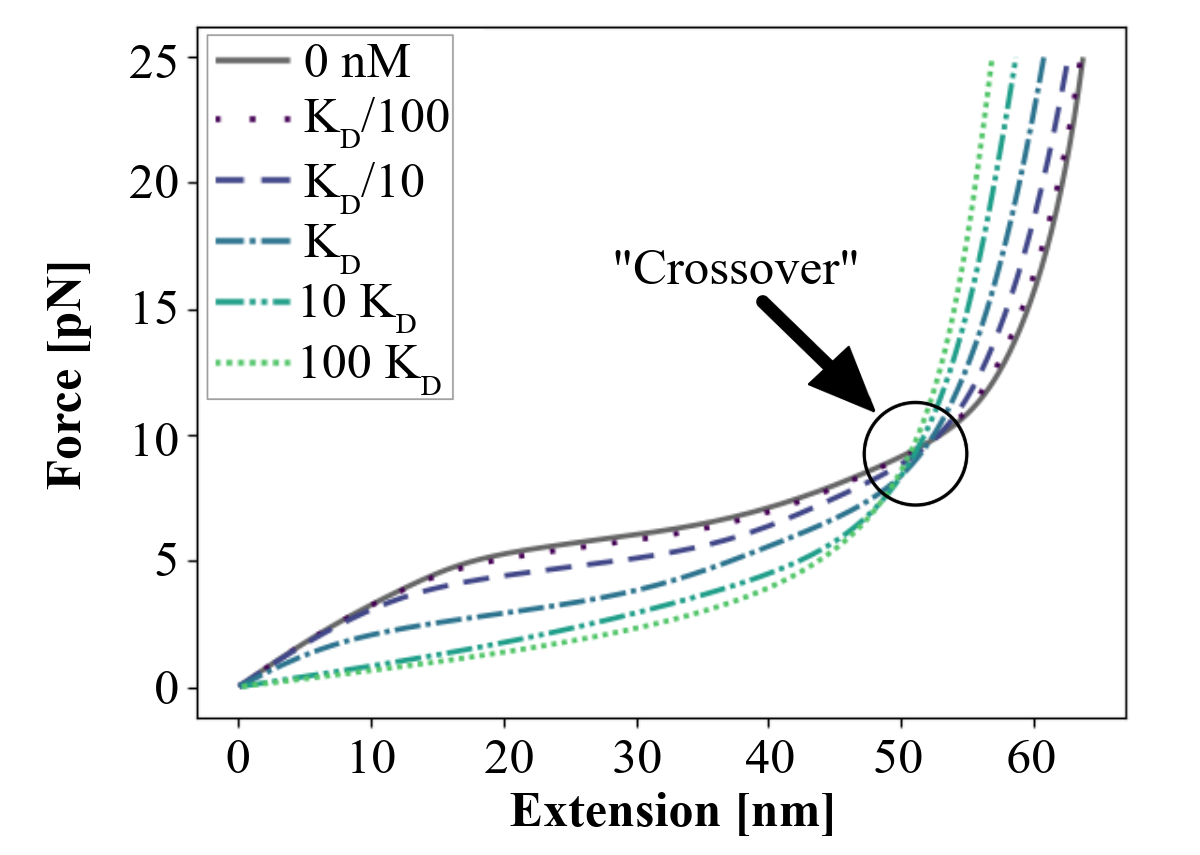}
	\caption{\label{crossdef} Definition and clear example of the characteristic crossover feature that is widely exhibited across concentration-dependent force extension simulations. Crossover points, highlighted here by a circle and pointed to with an arrow, are where multiple non-zero protein force extension curves cross the zero protein force extension curve.}
\end{figure}

\begin{figure}
	\centering
	\includegraphics[width=1.0\columnwidth]{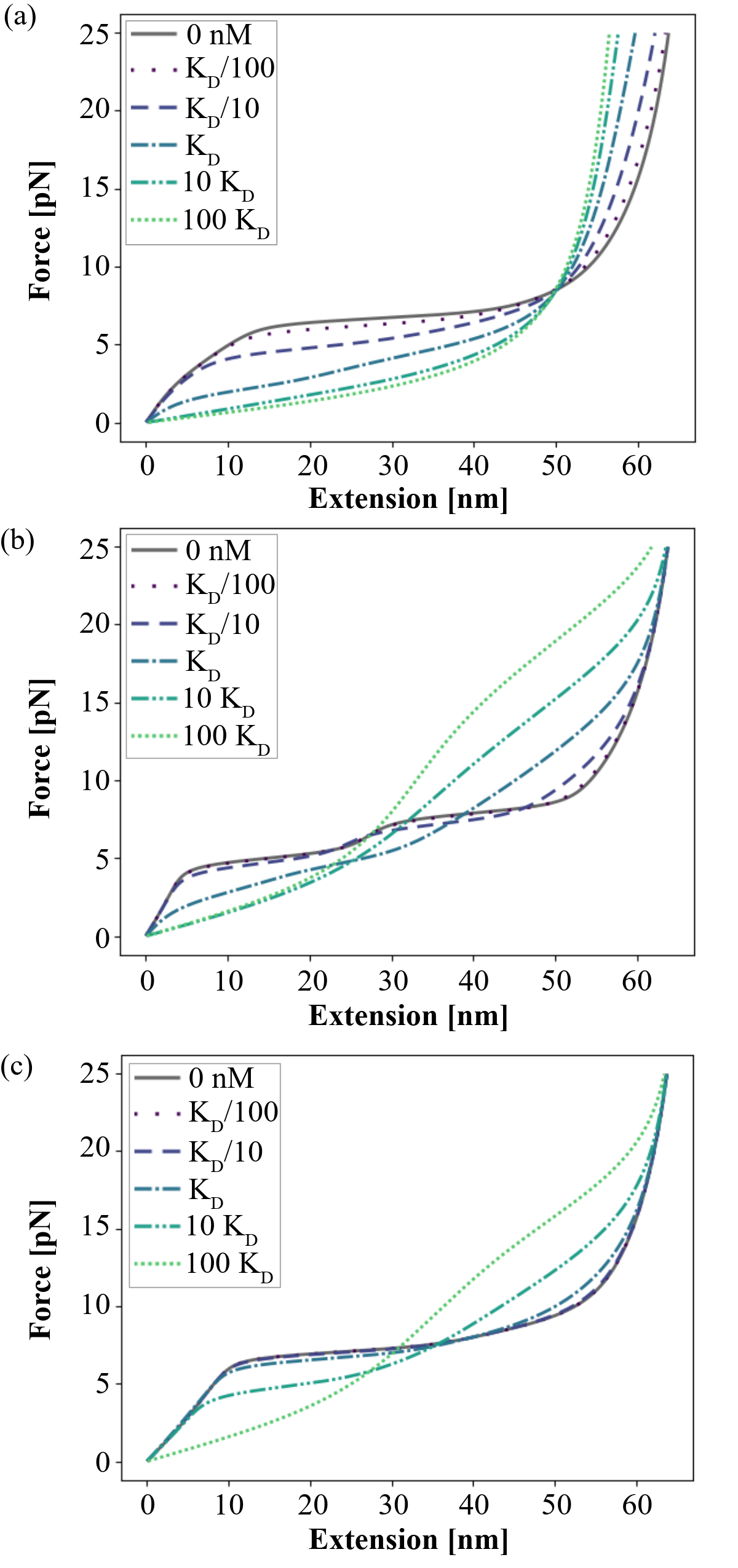}
	\caption{\label{crossex} Concentration-dependent force extension curves that exhibit (a) strong, (b) weak, and (c) marginal crossover convergence strength. (a) Strong crossover convergence is qualitatively described by crossover points having clear and distinct convergence upon visual inspection. (b) Weak crossover convergence qualitatively is when the individual crossing over of the zero protein curve for varying protein concentration force extension curves are clearly distinct and broadly distributed from one another. (c) Marginal crossover convergence is qualitatively described by a loosely discernible crossover point upon visual inspection though some minimal distribution of forces across protein concentration curves at the crossover point.}
\end{figure}

The concentration-dependent force extension curves resulting from these simulated RNA force spectroscopy experiments widely exhibit a prominent feature that we term ``crossover points". We define crossover points to be where multiple non-zero concentration force extension curves intersect (cross over) the corresponding zero protein force extension curve (Fig.~\ref{crossdef}). 
For some molecules, there exists multiple distinct crossover points (occurring each at different forces), but for all simulated molecules there is at least one crossover point. There is some degree of variation, however, to the convergence strength of these crossover points. For most crossover points, the crossovers for different protein concentrations converge and we see strong crossover convergence (Fig.~\ref{crossex}(a)). For some crossover points, though, the forces and extensions at which the force extension curves for different protein concentrations cross the zero-protein force extension curve exhibit a wider distribution; these points have weaker crossover convergence (Fig.~\ref{crossex}(b)).  We aim here to identify and measure the crossover points across the simulated RNA force spectroscopy experiments to discern the prevalence and strength of this crossover point feature.

To quantify this finding, we calculate the locations ($x_{cross}$, $F_{cross}$) and strengths $\zeta_{cross}$ of the crossover points for each of the 100 physiological RNA sequences for each of the three proteins HuR, U2AF2, and RBFOX1 (Appendix~\ref{sec:CalcCross}). Given the strengths and visual inspection of these crossover points, we determine a cutoff strength to be used for the classification of crossover points with strong convergence. A crossover point with a strength just at this cutoff upon visual inspection should appear marginal but acceptable in convergence strength (Fig.~\ref{crossex}(c)) and a large majority of crossover points with strength stronger then this cutoff should have clear convergence upon inspection. To ascertain a crossover strength cutoff, we surveyed the weakest accepted crossover point at various candidate cutoffs and chose the smallest crossover strength threshold for which the corresponding weakest accepted crossover point was still discernibly converged. The threshold identified for a crossover point to be classified as having strong convergence and therefore kept for analysis is $\zeta_{cross_{min}} = 2.5$ {pN}$^{-1}$.

Enforcing this minimum strength threshold $\zeta_{cross_{min}} = 2.5$ pN$^{-1}$, 85\% of the 300 total simulated concentration dependent force extension curves contain a strong crossover point. In more detail, 92\% of HuR simulations, 92\% of U2AF2 simulations, and 73\% of RBFOX1 simulations contain a strong crossover point. With these results, we conclude that strong crossover points are highly prevalent in concentration dependent force extension curves.\\

\subsection{\label{sec:Geometry} Protein binding domain length is extractable from crossover point location in predicted force extension curves\protect\\ }

\begin{figure}
	\centering
	\includegraphics[width=1.0\columnwidth]{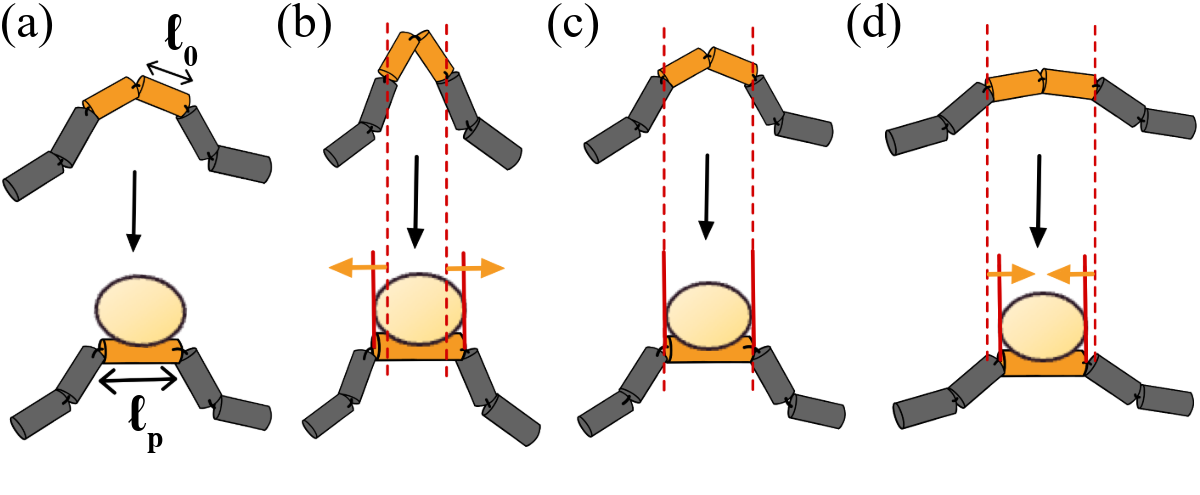}
	\caption{\label{forcereg} Illustration of how RNA extension is affected upon protein binding in varying force regimes and the connection between crossover points and protein binding domain lengths $\ell_p$. (a) In the absence of protein, each rod in the freely jointed chain model of RNA is of length $\ell_0$. Upon protein binding, the protein-bound segment of RNA is modeled by the freely jointed chain model as a single rod of length $\ell_p$. (b) In the low force regime, protein binding to RNA is expected to increase the end-to-end extension of the RNA molecule. (c) In the transition between the low and high force regimes, RNA extension is expected to be unchanged upon protein binding. In this case, RNA extension would ideally be independent of protein concentration and the concentration-dependent force extension curves would exhibit a strong crossover point. The end-to-end extension of the RNA molecule (and the external force) at this transition point is related to the length of the protein binding domain $\ell_p$. (d) In the high force regime, protein binding to RNA is expected to decrease the end-to-end extension of the RNA molecule.}
\end{figure}

Beyond their wide prevalence, concentration-dependent force extension curve crossover points also hold the potential to provide insight into protein binding domain geometries. When RNA is bound to protein, the end-to-end extension of the bound RNA segment is determined by the protein binding domain length $\ell_p$, also referred to here as the protein binding domain geometry. For low pulling forces, when unbound RNA is more likely to be in more dense spatial configurations, an unbound RNA segment tends to have smaller end-to-end extensions compared to when bound by the protein. Conversely, for high pulling forces, unbound RNA is more likely to be in taut or stretched configurations and an unbound RNA segment would have larger end-to-end extensions compared to when bound. It follows that at the transition between these two force regimes the RNA molecule would exhibit no change in extension upon protein binding, thus rendering the extension of the exterior part of the RNA independent of protein concentration and resulting in a crossover point (Fig.~\ref{forcereg}). As this crossover point is characterized by no change in extension upon protein binding, the extension of the RNA when this crossover point occurs should be related to the end-to-end extension of the RNA when bound to the protein - the protein binding domain length $\ell_p$. We further expect the force at which this crossover point occurs to also be related to the geometry of the binding domain $\ell_p$. We investigate here the dependence of crossover point forces on protein binding domain lengths and the possibility that binding domain lengths can be extracted from these proposed RNA force spectroscopy experiments.

To investigate the relationship between crossover point forces and protein binding domain lengths, we calculate the locations ($x_{cross}$, $F_{cross}$) and strengths $\zeta_{cross}$ of the crossover points for 1000 physiological RNA molecules for each of the three proteins HuR, RBFOX1, and U2AF2 (Appendix~\ref{sec:CalcCross}) at seven fixed protein binding domain lengths $\ell_{p,fix}$~$\in$~$\left\{10 \mathrm{\AA}, 15 \mathrm{\AA}, ..., 35 \mathrm{\AA}, 40 \mathrm{\AA}\right\}$. For reference, the physiological protein binding domain lengths for HuR, RBFOX1, and U2AF2 used elsewhere in this study are $\ell_{p,HuR} = 28.30 \mathrm{\AA}$ \cite{HuRRRM12PDBdoi,HuRPDBRRM12lit}, $\ell_{p,RBFOX1} = 23.27 \mathrm{\AA}$ \cite{RBFOX1PDBdoi,RBFOX1PDBlit}, and $\ell_{p,U2AF2} = 39.74 \mathrm{\AA}$ \cite{U2AF2PDBdoi,U2AF2PDBlit} and are obtained directly from their relevant RNA-bound protein structures found in the RCSB Protein Data Bank (PDB) (RCSB.org) \cite{PDB} (Appendix~\ref{sec:CalcPBDL}). As previously described, the crossover points that we are interested in when relating to binding domain geometry is the crossover point that marks the transition between the low and high force regimes. As such, the crossover point considered for this geometry analysis in the cases of the molecules where there are multiple crossover points is the point with the highest crossover force. These ``last" crossover points are further narrowed down to those with calculated strengths larger than the minimum strength threshold $\zeta_{cross_{min}} = 2.5$ pN$^{-1}$.

\begin{figure}
	\centering
	\includegraphics[width=1.0\columnwidth]{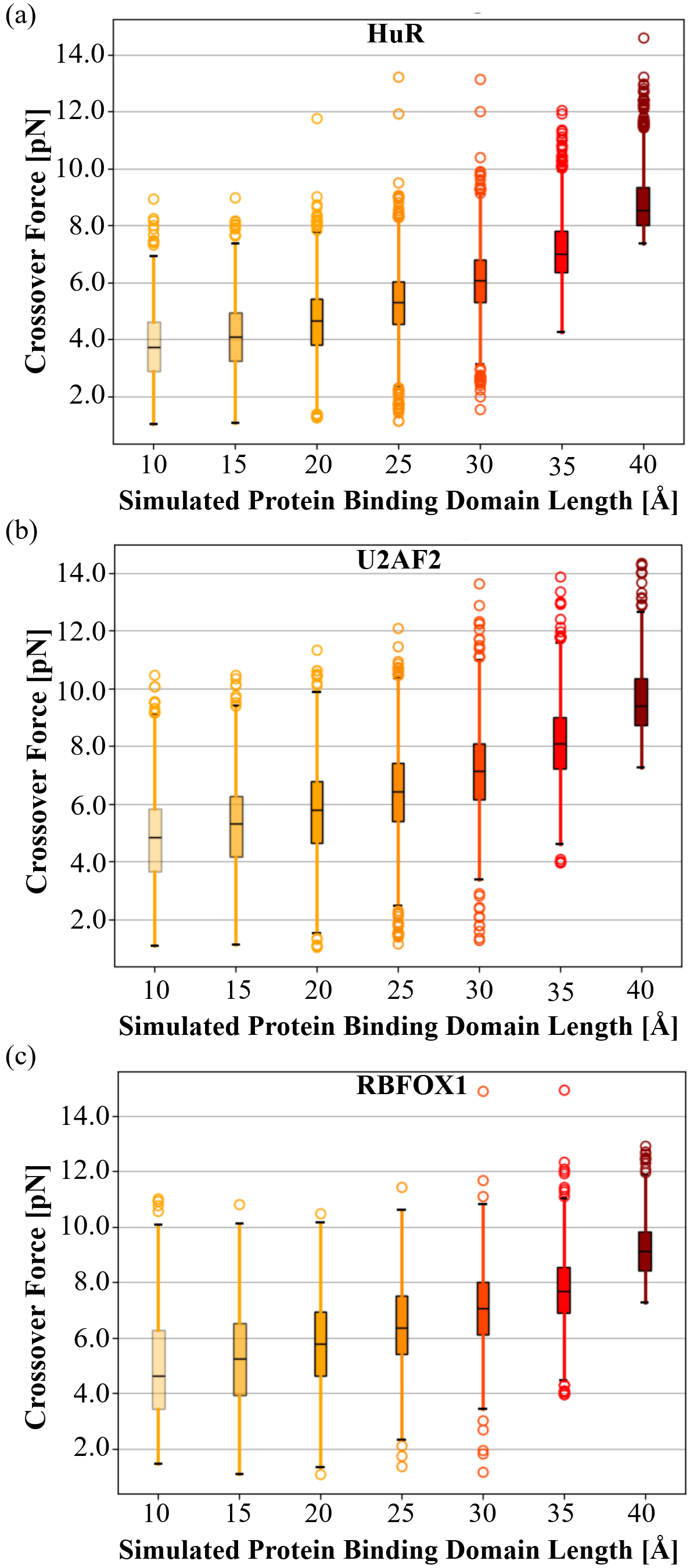}
	\caption{\label{crossglance} Distributions of crossover forces represented as box plots at increasing fixed protein binding domain lengths $\ell_{p,fix}$~$\in$~$\left\{10 \mathrm{\AA}, 15 \mathrm{\AA}, ..., 35 \mathrm{\AA}, 40 \mathrm{\AA}\right\}$ for (a) HuR, (b) U2AF2, and (c) RBFOX1. In each of (a), (b), and (c), the median crossover point force clearly increases with increasing fixed protein binding domain length, highlighting the positive association between binding domain length and crossover point force.}
\end{figure}

We aggregate last crossover points by protein and further by fixed protein binding domain lengths $\ell_{p, fix}$ to get isolated distributions of crossover point forces $F_{cross}$ for increasing sizes of protein binding domain geometries (Fig.~\ref{crossglance}). For each protein, we clearly see the median crossover force increases with larger fixed protein binding domain length. That we see median crossover forces increasing with larger protein geometries agrees with what we would expect from the concept of crossover points described above. At a crossover point, RNA does not change in extension upon protein binding, so for larger protein binding domains it would be necessary to apply a larger force and extend the RNA more for there to be no visible change in extension upon binding by the protein. With this result, we conclude that binding domain length is positively associated with crossover point force.

\begin{figure}
	\centering
	\includegraphics[width=1.0\columnwidth]{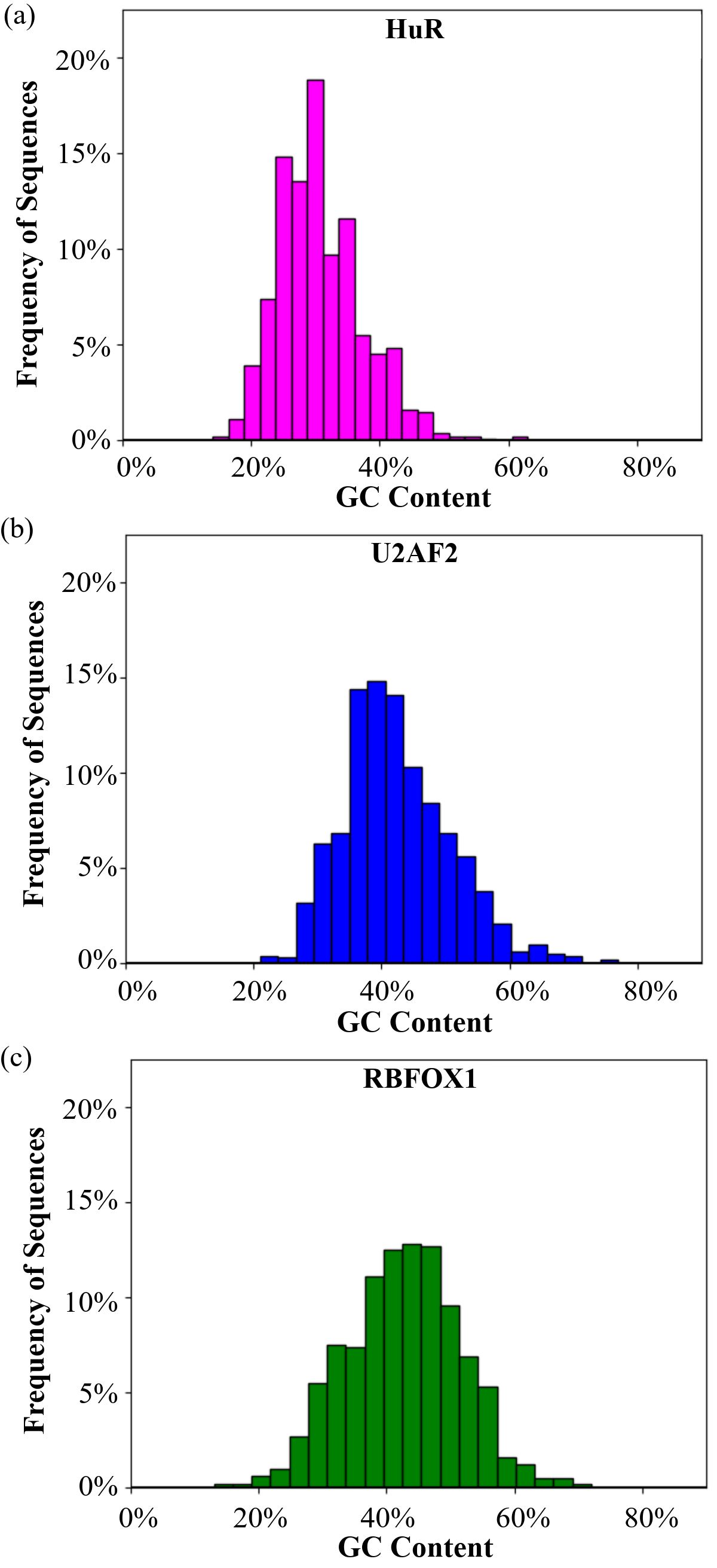}
	\caption{\label{GCcont} GC content distributions for the 1000 physiological sequences with natural binding sites corresponding to (a) HuR, (b) U2AF2, and (c) RBFOX1. From these GC content distributions, it becomes clear that the sequences used for each of these three simulated proteins have a broad range of GC content percentages, with RBFOX1 having a notably broader distribution. These broad GC content distributions point to the necessity of controlling for RNA structure strength when analyzing crossover point locations for binding domain geometry extraction.}
\end{figure}

To determine the possibility that protein binding domain lengths can be extracted from concentration dependent force extension curves using last crossover point forces, we expand upon the pervious analysis by also controlling for sequence content. In Fig.~\ref{crossglance}, we indeed see the positive association between median crossover force and binding domain length previously detailed, but it is also noted that the median crossover force for any given fixed protein binding domain length $\ell_{p,fix}$ varies across the three proteins. This is in part due to the effect of the strength of secondary structure on the location of the crossover point. For RNA with stronger base pairing and stacking effects, it would take more force to start to weaken the secondary structure and allow for interactions between secondary structure formation and protein binding. As we expect it to take larger pulling forces before we would start to see measurable protein binding for stronger secondary structures, we would also expect crossover forces to be affected by structure strength. We take a simplified measure of secondary structure strength, namely GC content. Generally speaking, we expect sequences with larger GC fractions to be capable of forming stronger secondary structures. Further, we do see differences in the distributions of GC content across the 1000 physiological sequences for the three proteins simulated here (Fig.~\ref{GCcont}), which could explain in part the protein-dependence of the median crossover forces seen in Fig.~\ref{crossglance}.

With these secondary structure strength distributions (Fig.~\ref{GCcont}) in mind, we additionally control the last crossover point force data for GC content. At this point, for each protein $p$, for each fixed protein binding domain length $d = \ell_{p,fix}$, and for each GC fraction bin $g$, we obtain last crossover force distributions, distribution means $\bar{F}_{cross, \left\{p, d, g\right\}}$, and the standard errors on the means $\sigma_{\bar{F}_{cross, \left\{p, d, g\right\}}}$. To obtain protein independent crossover data (for each GC fraction and for each fixed geometry) $\bar{F}_{cross, \left\{d, g\right\}}$, we take a weighted average of the mean crossover forces across the proteins. The uncertainty on these  weighted averages $\sigma_{\bar{F}_{cross, \left\{d, g\right\}}}$ are found by combining in quadrature the error due to uncertainties in the individual protein crossover force distributions with the error due to the differences between mean crossover forces for different proteins (Appendix~\ref{sec:CalcGeoData}).

\begin{figure}
	\centering
	\includegraphics[width=1.0\columnwidth]{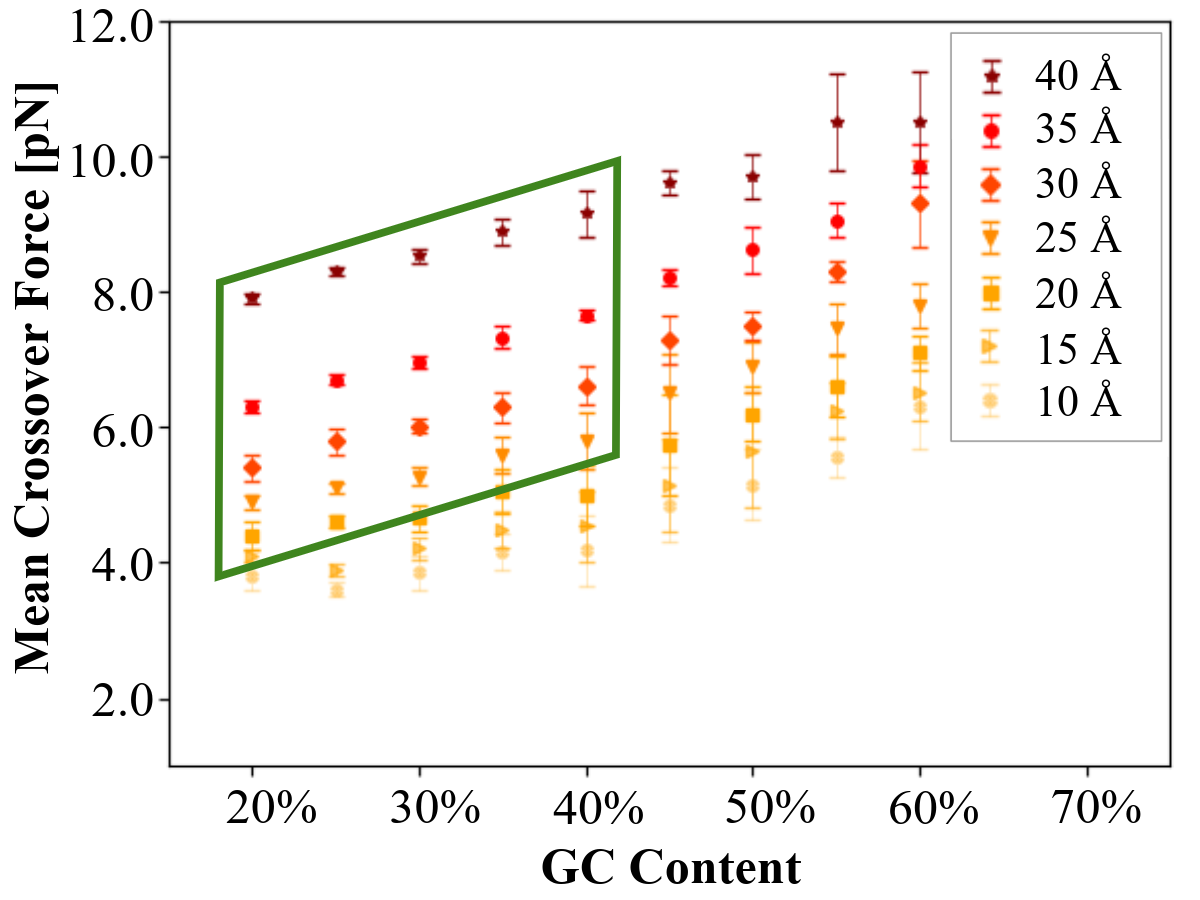}
	\caption{\label{geoextract} Scatter plot of mean crossover point force versus GC content at increasing fixed protein binding domain lengths $\ell_{p,fix}$~$\in$~$\left\{10 \mathrm{\AA}, 15 \mathrm{\AA}, ..., 35 \mathrm{\AA}, 40 \mathrm{\AA}\right\}$. Each point is the weighted average of the mean crossover forces for HuR, RBFOX1, and U2AF2 at the given fixed protein binding domain length $\ell_p$ and GC content fraction bin. When controlling for GC content, crossover point data are grouped in GC content bins with widths of 5\% GC content (i.e. 15\%-20\% GC content is the smallest group) according to the GC content of the corresponding sequence. The average crossover point force and its uncertainty for a given protein, geometry, and GC content is included in the corresponding weighted average calculation if there are 15 or more data points that can be averaged over. All 15\%-20\% GC content points and the 20\%-25\% GC content points for $\ell_{p,fix}$~$\in$~$\left\{10 \mathrm{\AA}, 15 \mathrm{\AA}, 20 \mathrm{\AA}, 35 \mathrm{\AA}\right\}$ come from just HuR data, the remaining  20\%-25\% GC content points come from HuR and RBFOX1 data, the 45\%-50\% GC content point at $\ell_{p,fix}$~$=$~$10 \mathrm{\AA}$ and all 50\%-55\% and 55\%-60\% GC content points come from RBFOX1 and U2AF2 data, and the rest of the points include data from all three proteins. The error bars represent the combination of both the uncertainties in the individual protein crossover force distributions and the deviations between the average crossover forces across the proteins. Small error bars therefore indicate both protein independence in the relationship between crossover point force and protein binding domain geometry (for a given GC content range) as well as precision in crossover point locations. The green box highlights the region (less than 40\% GC content and geometry larger than $25 \mathrm{\AA}$) where protein binding domain length can most accurately be extracted from crossover point location, as demonstrated by the decreased size of error bars and increased distance between points in this region.}
\end{figure}

We compare the protein cumulative (weighted) mean crossover force $\bar{F}_{cross, \left\{d, g\right\}}$ (and its total uncertainty $\sigma_{\bar{F}_{cross, \left\{d, g\right\}}}$) to GC content $g$ for each fixed protein binding domain length $d \equiv \ell_{p,fix}$ in Fig.~\ref{geoextract}. We see that the mean forces at which crossover points occur for different protein binding domain lengths are distinguishable from each other and protein independent when GC content is controlled for. This is most true in the case of smaller GC content and larger protein binding domain lengths, indicated by the box in Fig.~\ref{geoextract}, where the uncertainties are the smallest and the differences between crossover point forces for different geometries is the largest. We conclude that protein binding domain length (relative to binding footprint $N_{fp}$) can be determined from concentration dependent force extension curves given last crossover point force and GC content, especially for molecules with GC content less than 40\% and protein binding domain lengths larger than $25 \mathrm{\AA}$.\\

\section{\label{sec:Discussion}Discussion\protect\\ }

We have demonstrated through the implementation of a computational model of RNA and RNA binding protein interactions under an external force in the ViennaRNA program that single molecule force spectroscopy can be plausibly applied to measure the binding interactions between the RNA being pulled and free RNA binding proteins. Further, we have shown that by varying the free concentration of protein, there is a physically important crossover feature in the force extension curve data and that from this crossover point we can extract information on the geometry of the RNA-protein binding interaction. Single molecule force spectroscopy is an established method of measuring RNA structure \cite{OpticTweezerRef} and these results show that the applications of RNA force spectroscopy experiments can be expanded to also include the studies of interactions between RNA structure and RNA binding proteins as well as of the geometries of the domains where RNA binds to proteins. 

The RNA pulling simulations with interacting proteins presented in this paper would translate to the experiment we are proposing where RNA is pulled on in an optical or magnetic tweezer setup while it is in solution with an appropriate free concentration of protein. Ideally, as we have shown here, this experiment would be repeated with varying concentrations of protein around the measured effective binding constant. Larger changes in extension with increased protein concentration would indicate increased competition between RNA structure formation and protein binding. To experimentally measure the geometry of RNA when bound to the binding domain of the protein of interest as we propose here, we recommend these described experiments at varying protein concentrations to then be repeated for many different RNA sequences. The RNA sequences should be chosen such that RNA binding with the protein would occur, secondary structure will form (and compete with binding), and GC content of the sequences be controlled and for each sequence be less than 40\% of the primary structure. By repeating these experiments for many RNA sequences, and locating last crossover points for each RNA according to Appendix~\ref{sec:CalcCross}, the mean (last) crossover force found experimentally can be compared to Fig.~\ref{geoextract} to extract an approximate binding domain geometry.

During the course of this study, but not reported here, we also tried to discern the capability of the proposed RNA force spectroscopy experiment to measure effective binding constants. We attempted to extract the effective binding constants for HuR, RBFOX1, and U2AF2's physiological sequences by determining at which concentration the corresponding force extension curve was 50\% shifted from its 0nM curve to an over-saturated force extension curve. We found that this would not be a reliable application for this proposed experiment in that often the simulated force extension curve corresponding to the theoretical effective binding constant was within the simulated experimental noise and that this method was found to be relatively inaccurate beyond predicting theoretical effective binding constants' orders of magnitude. 

We also investigated the capability of this proposed experimental method to discern between binding modes of RNA binding proteins, such as modes with varying binding footprint sizes. In our effort, though, we found no significant difference in the crossover distributions for different footprint size binding modes and at this point believe the capability of RNA force spectroscopy to distinguish between binding modes of a given protein to be low.

There are some limitations to the model we have presented here that should be taken into account when interpreting our main results. One main limitation is the limited availability of absolute $N_{fp}$-mer binding affinities that are necessary for thermodynamic modeling of protein-RNA interactions. The sparsity of binding data limits the number of proteins that we are able to simulate here and with this method going forward. Another limitation of this methodology is the assumption of a system being in equilibrium. This thermodynamic equilibrium model is good enough for simulation purposes but should be taken into consideration when comparing simulated force-extension curves to experimentally determined ones. This is because most experiments do not occur quite so slowly that the system is in equilibrium, though some systems can be assumed to be near equilibrium when there is considerable agreement between the force extension curves when pulling on and when relaxing the RNA. The freely-jointed chain model described in Sec.~\ref{sec:ModelForce} is also an approximation that we make in our modeling that should be considered when comparing simulated force-extension curves to experiment. There are other models that can be similarly applied for the purpose of modeling RNA extension under an applied force, but the freely-jointed chain is sufficient for determining the plausibility of our proposed RNA force spectroscopy experiment and for studying features of these force extension simulations like crossover points. A further approximation that we make is that we simulate force-extension curves by incrementing the external force and computing the extension of the RNA (in equilibrium), as described in Sec.~\ref{sec:ModelForce}, and should be of extra consideration when comparing simulation to experiments where the extension of the RNA is controlled and the force measured. A final few limitations of the model as described include: protein-bound RNA is modeled as a fixed rod rather than a potentially flexible segment and tertiary structure elements, including pseudoknots, are explicitly excluded. These limitations and assumptions of the model we have presented here should be noted, especially when comparing these simulations to experiments quantitatively, but we believe that the model is sufficient to capture the essence of the proposed RNA spectroscopy experiment with interacting binding proteins for relevant RNA.

Some natural next directions considering the limitations of the model as described above include kinetic modeling of RNA and protein interactions (and under force), constant extension modeling of RNA-protein interactions while using an external force, and inclusion of some tertiary interactions in the model above. Kinetic modeling could be a particularly enlightening next direction because, as described previously, most experiments are not performed slowly enough for the system to be in equilibrium, and so modeling these interactions and this experiment outside of equilibrium would be particularly interesting. Constructing a constant extension model similar to the one described here would be similarly helpful when comparing simulation to experiment as force-extension curves are often generated using a constant-extension approach. Finally, including some tertiary interactions would be enlightening for some RNA for which tertiary interactions are exceptionally stabilizing and might not otherwise be described completely by this method, so this direction could be a worthwhile addition to this model in the future.

From this work, we conclude that single molecule force spectroscopy could be applied to study RNA-protein binding interactions and protein binding domain geometries. It is our aim that this study and future inspired studies enrich our fundamental understanding of how RNA and RNA-binding proteins interact.\\

\appendix

\section{\label{sec:Seqs} Determining natural RNA sequences for modeling protein binding\protect\\}

We obtain sequences with natural binding sites for U2AF2, RBFOX1, and HuR in similarly motivated approaches, but the detailed methods in each of the three cases diverge due to differences in data sources, RNA targets, etc. Thus, we will describe the methods for determining natural sequences for each of the three protein cases separately, starting with U2AF2, then RBFOX1, and finally HuR.

To obtain RNA sequences with naturally occurring binding sites for U2AF2, we start with genomic binding sites in the human genome identified by the Gene Yeo laboratory at UCSD via enhanced cross-linking and immunoprecipitation (eCLIP) \cite{U2AF2Data}. The U2AF2 binding site data collected by the Gene Yeo lab is available as a part of the ENCODE project \cite{Encode1,Encode2,Encode3}; we use the U2AF2 binding site data contained in the ENCFF290DFO peak file from the ENCSR893RAV experiment dataset. This peak file (ENCFF290DFO) contains the genomic locations of the U2AF2 binding sites (chromosome, start bp, stop bp, strand sense, etc.), which we use to locate 1000 randomly selected binding sites in the NCBI reference human genome \cite{humanNCBI, NCBIlit}. For these 1000 located binding sites, we extend the sequence symmetrically about the located binding site to obtain natural sequences 100 nucleotides in length. In the cases where the binding sites are near the start or end of the gene and cannot be extended symmetrically, we extend until we reach the start or end and extend in the other direction until a sequence of 100 nucleotides in length is obtained. For the sequences extended from binding sites identified on the negative sense strand, we ensure to replace them in the datasets with their reverse complements. In this way, we collect a dataset of 1000 natural U2AF2 binding sequences 100 nucleotides in length. In the analyses where 100 natural sequences are needed rather than 1000, we use the first 100 sequences of this randomly selected dataset of 1000 sequences.

The RBFOX1 natural sequences are obtained using a similar process to U2AF2 described above. The RBFOX1 genomic binding sites were identified with CLIP using mouse whole-brain tissue by Weyn-Vanhentenryck \emph{et al.} \cite{RBFOX1Data}. The genomic locations of the RBFOX1 binding site clusters is available in Weyn-Vanhentenryck \emph{et al.}'s \cite{RBFOX1Data} supplementary material, which contains information such as genomic location, cluster length, signal strength, strand sense, and CLIP tag count. As there are CLIP clusters of varying strength in this dataset, we take care to refine the RBFOX1 data we use to clusters less than 100 nucleotides in length and to those within the top ~1\% and ~10\% of RBFOX1 CLIP tags by number ($>$10 and $>$5 CLIP tags, respectively). There are 196 RBFOX1 binding sites in the top ~1\% and 1812 RBFOX1 binding sites in the top ~10\%. We randomly select 100 of the top 196 binding sites and 1000 of the top 1812 binding sites and use their genomic locations as described in the supplementary detail of Weyn-Vanhentenryck \emph{et al.} \cite{RBFOX1Data} to locate 100 and 1000 binding sites in the NCBI reference mouse genome \cite{mouseNCBI, NCBIlit}. At this point, we extend the 100 and 1000 RBFOX1 sequences just as we did for U2AF2 (using the reference mouse genome rather than human) and similarly calculate the reverse complements for sequences with binding sites found on the negative sense strand. From this method, we obtain datasets of 100 and 1000 sequences that are 100 nucleotides in length containing natural RBFOX1 binding sites.

The HuR natural sequences were collected slightly differently from those for U2AF2 and RBFOX1 for the reasons that HuR's binding sites are located in the transcriptome rather than the genome and the HuR binding site data that we use here contain binding site sequences in addition to locations. HuR binding sites are located in the transcriptome rather than the genome due to HuR's widely recognized activity in untranslated regions of mature mRNA \cite{HuRBackground}, while U2AF2 and RBFOX1 act on unprocessed pre-mRNA. HuR binding sites were identified via cross-linking and immunoprecipitation (CLIP) and summarized by binding site sequence, transcript id, start nucleotide, stop nucleotide, etc. by Kishore \emph{et al.} \cite{HuRData}. For 1000 randomly selected binding sites from this dataset, we locate their locations in reference human transcriptomes. Notably, some of the binding sites identified by Kishore \emph{et al.} were reported with locations (transcript id) that correspond to data contained in the European Nucleotide Archive (ENA) (ebi.ac.uk/ena/) \cite{ENAlit} and some binding sites had reported locations (transcript id, start nucleotide, stop nucleotide) that matched an older version of the NCBI reference human transcriptome \cite{NCBIlit}, rendering the start and stop nucleotides generally inaccurate for current references. Thus, to locate the 1000 randomly selected binding sites in the human transcriptome, binding sites with transcript identifiers of the scheme ``XX\_\#.." were located in the corresponding current NCBI reference transcript \cite{humanNCBI, NCBIlit} by its sequence and binding sites with transcript identifiers of the scheme ``XX\#.." were located also by its sequence in the corresponding current ENA reference transcript using the browser feature (ebi.ac.uk/ena/browser/api/fasta/XX\#..) \cite{ENAlit}. Then, we extend the 1000 HuR sequences just as we did for U2AF2 (using the reference human transcripts rather than genomes). At this point, we obtain 1000 sequences containing natural HuR binding sites that are 100 nucleotides in length. In the analyses where 100 natural sequences are needed rather than 1000, we use the first 100 sequences of this randomly selected dataset of 1000 sequences.\\

\section{\label{sec:FirstDistin}Determining protein concentration when predicted extension becomes distinguishable from RNA-only force extension curve\protect\\}

After calculating for each physiological sequence the average area under the generated noisy force-extension curves and the standard deviation on the noisy area, a search for the smallest corresponding protein concentration $c$ at which the resulting force extension curve deviates by more than 2 standard deviations from the average noisy area starts at $c=0.1$ nM. If the area does not deviate by more than 2 standard deviations from the noise, the protein concentration is incremented by a factor of 2 and this evaluation process repeated until either a first distinguishable concentration is determined or until protein binding is deemed uncapable of being distinguished at any concentration. The criteria for when protein binding for a given sequence is determined to be indistinguishable is that (1) the area under the predicted force extension curve has for some nonzero protein concentration deviated from the area under the predicted zero protein force extension curve by more than 0.1\% (to ensure the search is not terminated prematurely), (2) the predicted force extension curve has not for any concentration deviated by more than two standard deviations from the noise distributed about the zero protein force extension curve, and (3) the incrementing of protein concentration by a factor of 2 ceases to have an effect on the predicted force extension curves as measured by the deviation from the zero concentration force extension curve area for two consecutively simulated protein concentrations being less than 0.1\% different (indicating saturation). This search process was repeated separately for maximum forces of 5, 6, 7, 8, 9, and 10 pN. We found that 5 pN was too small of an upper limit and 10pN too large while 6, 7, 8, and 9 pN were mostly comparable to each other in regards to fractions of RNA molecules eventually found distinguishable from noise and the distributions of protein concentrations (relative to corresponding effective binding constants) when first distinguishable from noise. Thus, we here only report results for a maximum force of 7 pN.\\

\section{\label{sec:CalcCross}Calculating location and strength of crossover points\protect\\}

For a given sequence and protein, the crossover points are located by first calculating force and extension data for six different free concentrations c of the given protein, given its predicted effective binding constant $K_D$ at zero force: 0 nM, $0.01 \cdot K_D$, $0.1 \cdot K_D$, $K_D$, $10 \cdot K_D$, and $100 \cdot K_D$. The discrete force and extension data are then interpolated using a cubic spline to obtain continuous, predicted force-extension curves $F_{c}(x)$ for each free protein concentration $c$. For each non-zero concentration, we locate individual crossover points by finding each $x_{cross_{ind}}$ that locally minimizes $[F_{c}(x_{cross_{ind}}) - F_{0}(x_{cross_{ind}})]^2$. Specifically, approximate individual crossover point location bounds are identified by changes in sign in $\Delta F(x) = F_{c}(x) - F_{0}(x)$ on steps of $\Delta x = 0.01$ nm; a lower bound $x_{min,i}$ is indicated by $x$ for which $\Delta F(x) \cdot \Delta F(x + \Delta x) < 0 $. The squared difference $[F_{c}(x_{cross_{ind}}) - F_{0}(x_{cross_{ind}})]^2$ is minimized using the scipy.optimize.minimize function in python (which by default uses the Broyden–Fletcher–Goldfarb–Shanno (BFGS) optimization method) on each set of bounds $x_{cross_{ind},i} \in [x_{min, i}, x_{min, i}+ \Delta x)$, with the minimizer initialized at $x_{min, i} + 0.5 \cdot \Delta x$, to find a list of individual crossover points {$x_{cross_{ind},i}$} across each concentration. This search for individual crossover points occurs over the range of $x \in [ x_{min} , x_{max}]$, where $x_{min}$ is the smallest extension calculated among all six pulling simulations and $x_{max}$ is the extension corresponding to the force $F_{max}$ at which the probability of the protein being bound becomes mostly negligible. This maximum force $F_{max}$ is determined according to when the probability for the very last protein to be bound to the strongest binding site $P_{bound}$ becomes less than 0.01 for the given concentration $c$. Here, the probability to be bound is computed as
\begin{equation}
	P_{bound}(F, c) = \frac{Z_{bound}(F,c)}{Z_{bound}(F,c) + Z_{unbound}(F)}
\end{equation}
with
\begin{equation}
	Z_{unbound}(F) = [\frac{k_B T}{F \ell_0} \sinh (\frac{F \ell_0}{k_B T})]^{N_{fp} \cdot (\frac{x_b}{\ell_0})}
\end{equation}
and
\begin{equation}
	Z_{bound}(F, c) = \frac{c}{K_D} \frac{k_B T}{F \ell_p} \sinh (\frac{F \ell_p}{k_B T}),
\end{equation}
where $K_D$ is the absolute binding affinity of the protein to the strongest binding site in the given sequence as determined from calibrated RNAcompete data. In practice, the solution for $F_{max}$ is determined in logarithmic space.

After collecting individual crossovers $x_{cross_{ind}}$ across all concentrations, the final crossover points are determined by finding which unique extensions $x_{cross}$ minimize $\sum_{c \neq 0} [F_{c}(x_{cross}) - F_{0}(x_{cross})]^2$, using each individual crossover $x_{cross_{ind}}$ as an initial position for the BFGS minimizer. The unique crossover extensions $x_{cross}$ found this way are then converted to average crossover forces $F_{cross}$ and strengths $\zeta_{cross}$ by taking the average $F_{cross} = \frac{1}{N_{c \neq 0}} \sum_{c \neq 0} F_c(x_{cross})$ and the inverse of the standard deviation $\zeta_{cross} = \left(\sqrt{\frac{1}{N_{c \neq 0}-1} \sum_{c \neq 0} [F_{c}(x_{cross}) - F_{cross}]^2}\right)^{-1}$. 

\begin{figure}
	\centering
	\includegraphics[width=1.0\columnwidth]{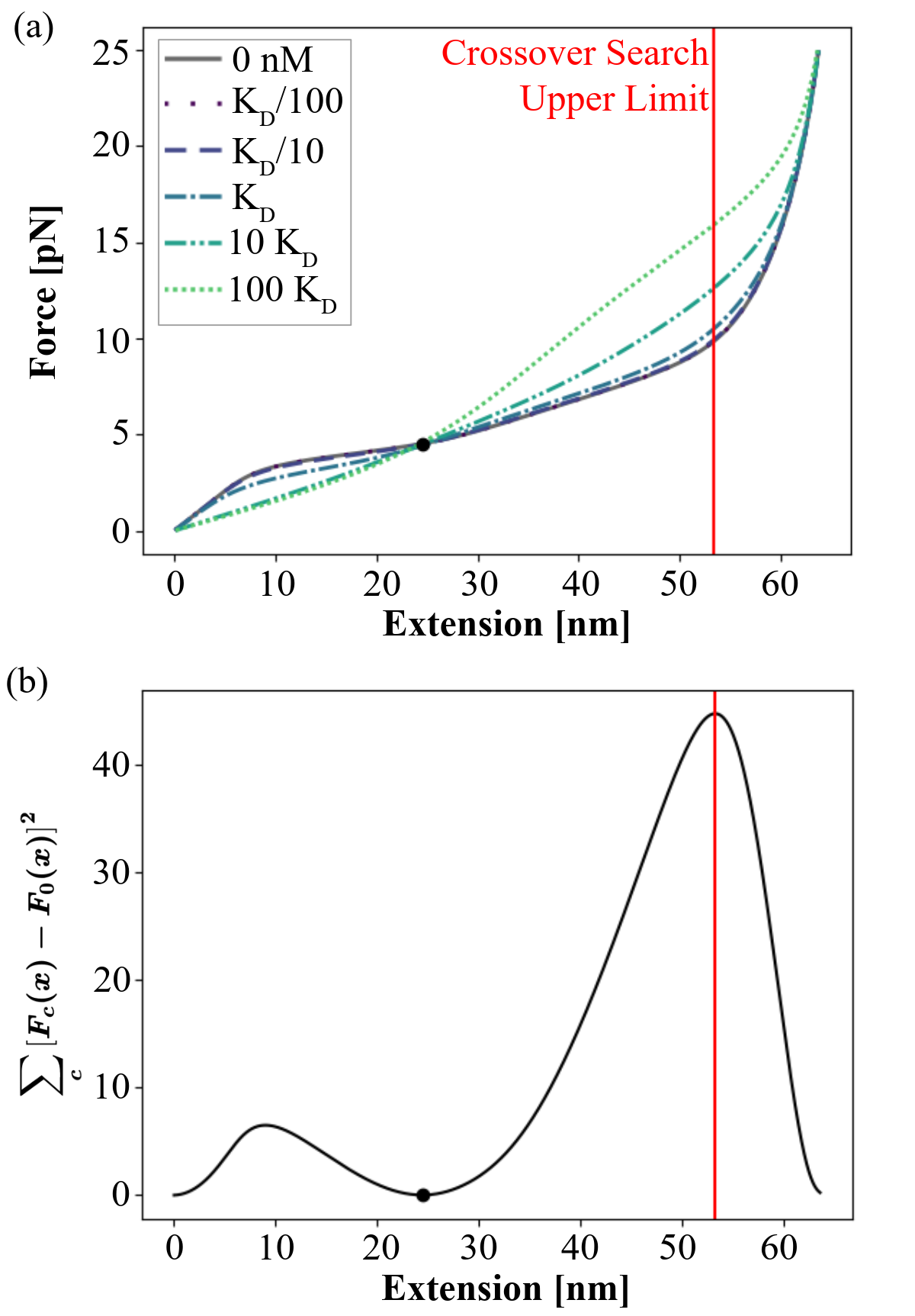}
	\caption{\label{sumfluc} (a) Example concentration-dependent force extension curve with the upper extension limit for the crossover point search $x_{limit}$ indicated by a vertical red line. This upper extension limit $x_{limit}$ corresponds to the first maximum from the right in the (b) totaled force fluctuation function $\sum_{c \neq 0} [F_{c}(x_{cross}) - F_{0}(x_{cross})]^2$, where the minima of this function correspond to prospective crossover point(s) (indicated by black dots). The use for this upper extension limit $x_{limit}$ in the search for crossover point locations is specifically to correct for the cases where the minimizer for $\sum_{c \neq 0} [F_{c}(x_{cross}) - F_{0}(x_{cross})]^2$ is initiated above this limit and illegitimate crossover points are incorrectly identified.}
\end{figure}

The crossover points are then narrowed down to those with $F_{cross} \geq 1$~pN and $x_{cross} < x_{limit}$, where $x_{limit}$ is the extension that limits from below the region where the minimizer, if placed wherein, would progress to an extension at which all of the proteins have unbound from the RNA. Specifically, in the landscape being minimized $\sum_{c \neq 0} [F_{c}(x_{cross}) - F_{0}(x_{cross})]^2$, the list of crossover points $x_{cross}$ are the local minima and the upper extension limit $x_{limit}$ is the first maximum from the right (Fig.~\ref{sumfluc}) as determined by directly finding the first extension coming from the right for which the function $\sum_{c \neq 0} [F_{c}(x_{cross}) - F_{0}(x_{cross})]^2$ begins to decrease and has value above a threshold $0.5$~pN$^2$ (which addresses occasional noise at larger extensions for smaller protein binding domain geometries).
\\

\section{\label{sec:CalcPBDL}Determining physiological protein binding domain length\protect\\}

The protein binding domain lengths for RBFOX1, HuR, and U2AF2 were determined from the RCSB Protein Data Bank (PDB) Mol* 3D viewer (rcsb.org/3d-view) \cite{PDB,3Dview} using relevant crystal structures of each of the proteins bound to RNA. 

The bound crystal structure for RBFOX1 (PDB ID: 2ERR)\cite{RBFOX1PDBdoi, RBFOX1PDBlit} shows the protein bound to an RNA segment 7 nucleotides in length \cite{3Dview,PDB}. As this 7 nucleotide segment length is the footprint size used in our model and in the normalized RNAcompete data, we use the PDB Mol* 3D viewer measurement tool  \cite{PDB,3Dview} to determine the distance between the first and last bases of the bound RNA exactly as imaged and then add to this measurement the distance between ssRNA bases $x_b = 7 \mathrm{\AA}$ to account for the additional backbone extending beyond the first and last bound bases (as the measurement tool only measures distance between bases). From this crystal structure measurement and the backbone correction, we approximate the physiological protein binding domain length for RBFOX1 to be $\ell_{p,RBFOX1} = 16.27 \mathrm{\AA} + 7.00 \mathrm{\AA} = 23.27 \mathrm{\AA}$.

The bound crystal structure for HuR (PDB ID: 4ED5) depicts the two N-terminal tandem RNA-recognition motifs RRM1/2 bound to an 8 nucleotide long RNA segment (labeled U3 to U10) and the 5' phosphate of a ninth nucleotide (U11) \cite{HuRRRM12PDBdoi,HuRPDBRRM12lit,3Dview,PDB}. The choice of the crystal structure of the tandem RRM1/2 binding domains over that of HuR's RRM3 binding domain complexed with RNA is due to the binding activity of the RRM1/2 domains with AU-rich RNA, whereas the RRM3 has been noted to be more active in binding poly-A tails \cite{HuRPDBRRM12lit}. While the crystal structure shows HuR complexed with RNA 9 nucleotides in length, the footprint size used in the RNAcompete data and thus the model here is 7 nucleotides and so we need to make a choice of which 7 continuous nucleotides of the 9 imaged to approximate the binding footprint and to measure the end-to-end distance. As the U11 nucleotide is not completely complexed with HuR and as RRM1 (complexed with U5-U8 and U10) has been found to be the primary AU-rich RNA binding domain of the two tandem binding domains \cite{HuRPDBRRM12lit}, we make the choice to focus on the 7 nucleotide RNA segment U4-U10 (forgoing U3 and U11). Using the end-to-end distance measurement between the U4 and U10 nucleotides, and again correcting for the additional backbone extending beyond the first and last bound bases (in our approximation) which is unaccounted for by the base-to-base PDB Mol* 3D viewer measurement tool  \cite{PDB,3Dview}, we approximate the physiological protein binding domain length for HuR to be $\ell_{p,HuR} = 21.30 \mathrm{\AA} + 7.00 \mathrm{\AA} = 28.30 \mathrm{\AA}$.

The crystal structure for U2AF2 shows the protein bound to an RNA molecule (\textit{AdML}-C5 oligonucleotide) that is 8 nucleotides in length, labeled U2-C9 (PDB ID: 7S3A) \cite{U2AF2PDBdoi, U2AF2PDBlit,3Dview,PDB}. The choice of the U2AF2-RNA complex with the C5 \textit{AdML} Py tract variant over the G5 and A5 variants is not so significant as the difference in overall structures of the bound RNA variants is very minimal \cite{U2AF2PDBlit}. Since the footprint size used in our model and in the normalized RNAcompete data is 7 nucleotides in length, we have the ability to use the PDB Mol* 3D viewer measurement tool  \cite{PDB,3Dview} to determine the distance either between the U2 and C8 nucleotides or between the U3 and C9 nucleotides. Upon examination, the U2 nucleotide appears more bound in the structure than the C9 nucleotide and the difference between the two distance measurements is sufficiently small ($\approx 1.3 \mathrm{\AA}$) that it is acceptable to use the distance measurement between the U2 and C8 nucleotides to determine the geometry. Using this distance measurement from the crystal structure, and again correcting for the additional backbone extending beyond the first and last bound bases unaccounted for by the base-to-base measurement tool, we approximate the physiological protein binding domain length for U2AF2 to be $\ell_{p,U2AF2} = 32.74 \mathrm{\AA} + 7.00 \mathrm{\AA} = 39.74 \mathrm{\AA}$.\\

\section{\label{sec:CalcGeoData} Calculating protein cumulative crossover forces for geometry extraction analysis\protect\\}

To obtain protein cumulative crossover force data for each fixed protein binding domain length $d \equiv \ell_{p,fix}$ and for each GC fraction $g$, given mean last crossover forces $\bar{F}_{cross, \left\{p, d, g\right\}}$ and standard errors on those means $\sigma_{\bar{F}_{cross, \left\{p, d, g\right\}}}$ for each protein $p \in$ \{HuR, RBFOX1, U2AF2\} with at least $15$ relevant crossover data points, we take a weighted average across the proteins 
\begin{equation}
	\bar{F}_{cross, \left\{d, g\right\}} = \frac{\sum_p \bar{F}_{cross, \left\{p, d, g\right\}} \cdot \sfrac{1}{\sigma_{\bar{F}_{cross, \left\{p, d, g\right\}}}^2}}{\sum_p \sfrac{1}{\sigma_{\bar{F}_{cross, \left\{p, d, g\right\}}}^2}}.
\end{equation}

To get the total uncertainty on this weighted average $\sigma_{\bar{F}_{cross, \left\{d, g\right\}}}$, we combine in quadrature the error due to variation in the last crossover forces for each protein 
\begin{equation}
	\delta_{1, \bar{F}_{cross, \left\{d, g\right\}}} = \left(\sqrt{\sum_{p} \frac{1}{\sigma_{\bar{F}_{cross, \left\{p, d, g\right\}}}^2}}\right)^{-1}
\end{equation}
with the error due to the difference in mean last crossover forces between each protein
\begin{equation}
	\delta_{2, \bar{F}_{cross, \left\{d, g\right\}}} = \sqrt{\frac{1}{N-1} \sum_p \left( \bar{F}_{cross, \left\{p, d, g\right\}} - \bar{F}_{cross, \left\{d, g\right\}} \right)^2},
\end{equation}
where $N$ is the number of proteins $p$ with sufficient data for that GC fraction $g$, and arrive at

\begin{equation}
	\sigma_{\bar{F}_{cross, \left\{d, g\right\}}} = \sqrt{\left( \delta_{1, \bar{F}_{cross, \left\{d, g\right\}}}\right)^2 + \left(\delta_{2, \bar{F}_{cross, \left\{d, g\right\}}}\right)^2 }.
\end{equation}
\\


\bibliographystyle{unsrt_mod}
\bibliography{biblio}

\end{document}